\documentclass[11pt]{article}

\usepackage{amsmath,amsthm} 
\usepackage{amssymb,mathrsfs} 
\usepackage{a4wide} 
\usepackage{graphicx} 
\usepackage{color}
\usepackage{enumerate}  
\usepackage{fullpage}
\usepackage{epsfig}
\usepackage{caption}
\usepackage{subcaption}
\usepackage{dsfont}

\newtheorem{remark}{\bf Remark}[section]
\usepackage{dsfont}
\usepackage{float}
\usepackage{xcolor}
\usepackage{subcaption}

\newcommand{\R}{\mathbb{R}}
\newcommand{\E}{\mathbb{E}}

\newcommand{\de}{\Delta}
\newcommand{\n}{\nabla}

\newcommand{\Lop}{\mathcal{L}}

\newcommand{\set}[1]{\mathbb{#1}}        

\newcommand{\e}{\varepsilon}

\newcommand{\vp}{A}
\newcommand{\op}[1]{\mathcal{#1} }

\newcommand{\du}{\partial}

\begin{document}

\title{Local and Global Perspectives on Diffusion Maps in the Analysis of Molecular Systems}

\author{Z. Trstanova$^{1}$, B. Leimkuhler$^{1}$ and T. Leli\`evre$^{2}$ \\
	{\small $^{1}$School of Mathematics, University of Edinburgh, James Clerk Maxwell Building, } \\
	{\small Peter Guthrie Tait Road, Edinburgh EH9 3FD, UK} \\
	{\small $^{2}$  CERMICS (ENPC), INRIA, 77455 Marne-la-Vall\'ee, France}
}		

\maketitle

\begin{abstract}

Diffusion maps approximate the generator of Langevin dynamics from simulation data.  They afford a means of identifying the slowly-evolving principal modes of high-dimensional molecular systems.  When combined with a biasing mechanism, diffusion maps can accelerate the sampling of the stationary Boltzmann-Gibbs distribution. 
In this work, we contrast the local and global perspectives on diffusion maps, based on whether or not the data distribution has been fully explored.  In the global setting, we use diffusion maps to identify metastable sets and to approximate the corresponding committor functions of transitions between them. We also discuss the use of diffusion maps {\em within} the metastable sets, formalising the locality via the concept of the quasi-stationary distribution and justifying the convergence of diffusion maps within a local equilibrium. This perspective allows us to propose an enhanced sampling algorithm. We demonstrate the practical 
relevance of these approaches  both for simple models and for molecular dynamics problems (alanine dipeptide and deca-alanine). 
\end{abstract}

\maketitle

\section{Introduction}

The calculation of thermodynamic averages for complex models remains a fundamental challenge in computational chemistry~\cite{chem}, materials modelling~\cite{bereau2016research} and biology~\cite{shaffer2016enhanced}. In typical situations, the potential energy $U$ of the system is known, and the states are assumed to be distributed according to the Boltzmann-Gibbs distribution with density
\[
\rho_{\beta} =Z^{-1} e^{-\beta U}.
\]
The difficulty arises due to the high-dimensional, multimodal nature of the target distribution in combination with limited computational resources. The multimodality of the target distribution causes that the high-likelihood conformational states are separated by low-probability transition regions.  In such a setting, the transitions between critical states become ``rare events'', meaning that naive computational approaches converge slowly (if at all) or produce inaccurate results.
Moreover, standard MD simulations of large proteins (>500 residues) run on regular CPUs allow to
simulate dynamics over hundreds of ns (or a few $\mu$m max) in a reasonable amount of time. This
time scale is often not enough to sample relevant conformational transitions which may occur on
the ms time scale. 

\begin{figure}[!h]

    \centering
    \begin{subfigure}[b]{0.3\textwidth}
        \includegraphics[width=\textwidth]{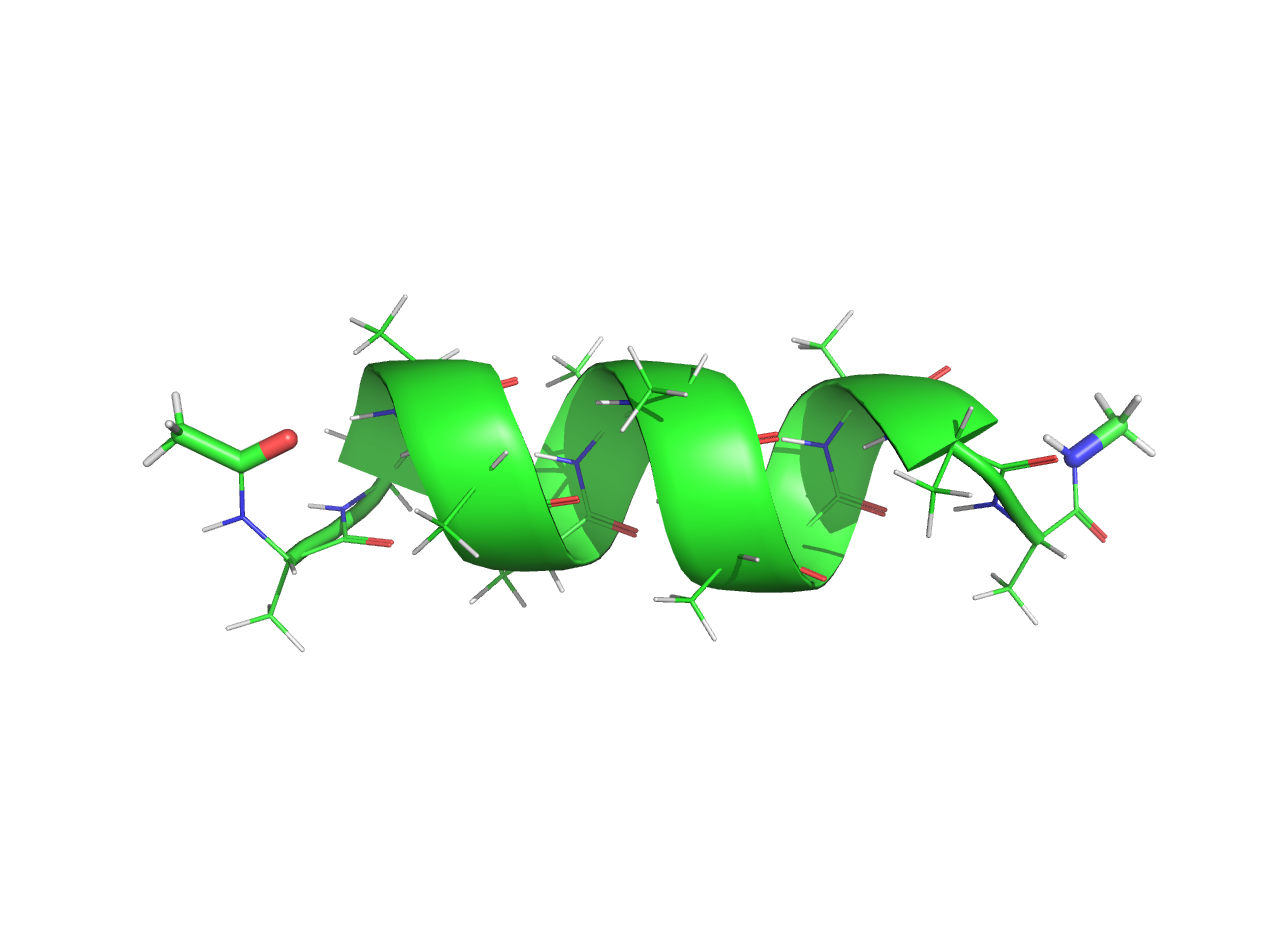}
                         \caption{A}
         \end{subfigure}
~
    \begin{subfigure}[b]{0.3\textwidth}
        \includegraphics[width=\textwidth]{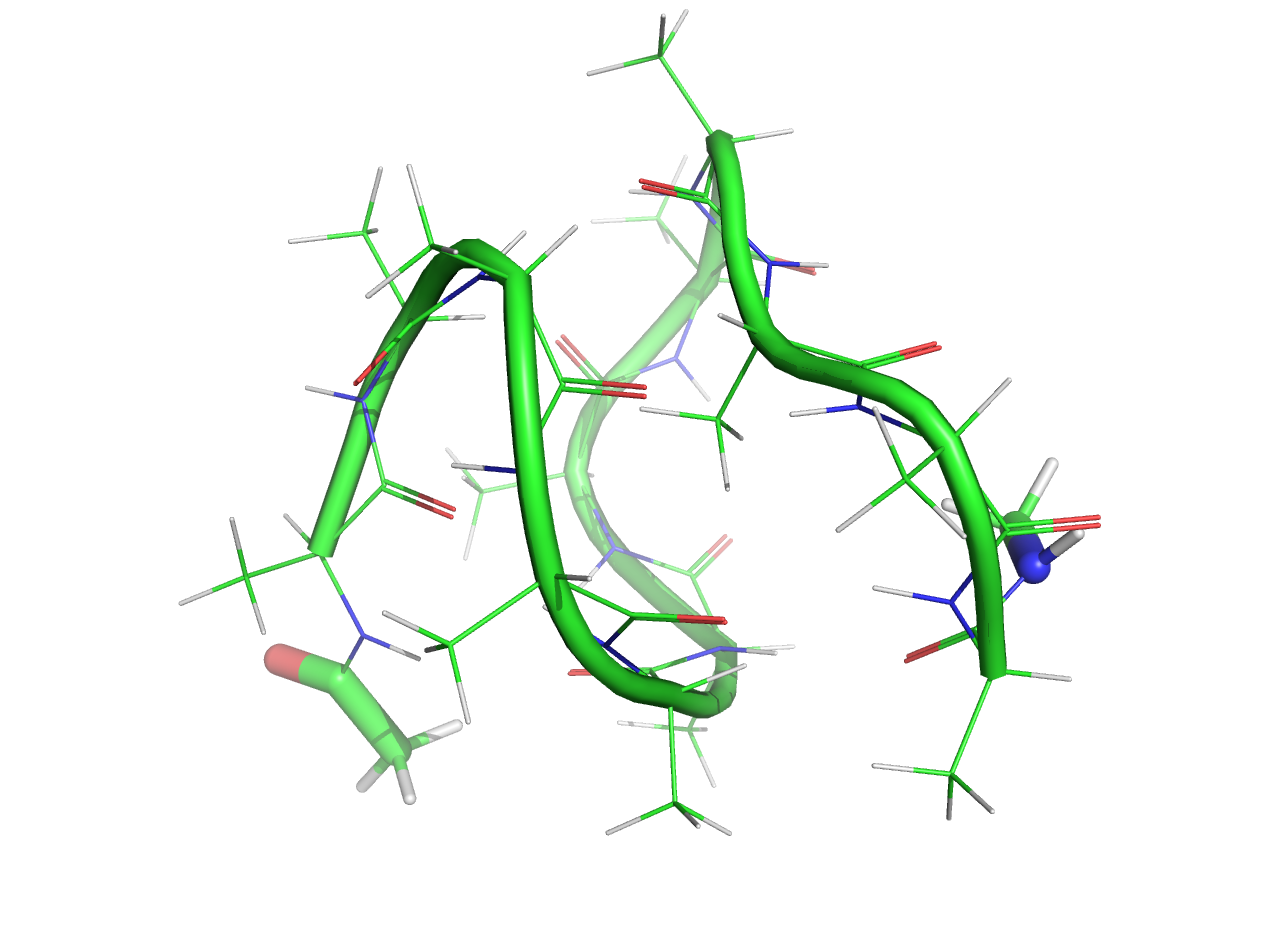}
                         \caption{B1}
         \end{subfigure}
        ~
          \begin{subfigure}[b]{0.3\textwidth}
        \includegraphics[width=\textwidth]{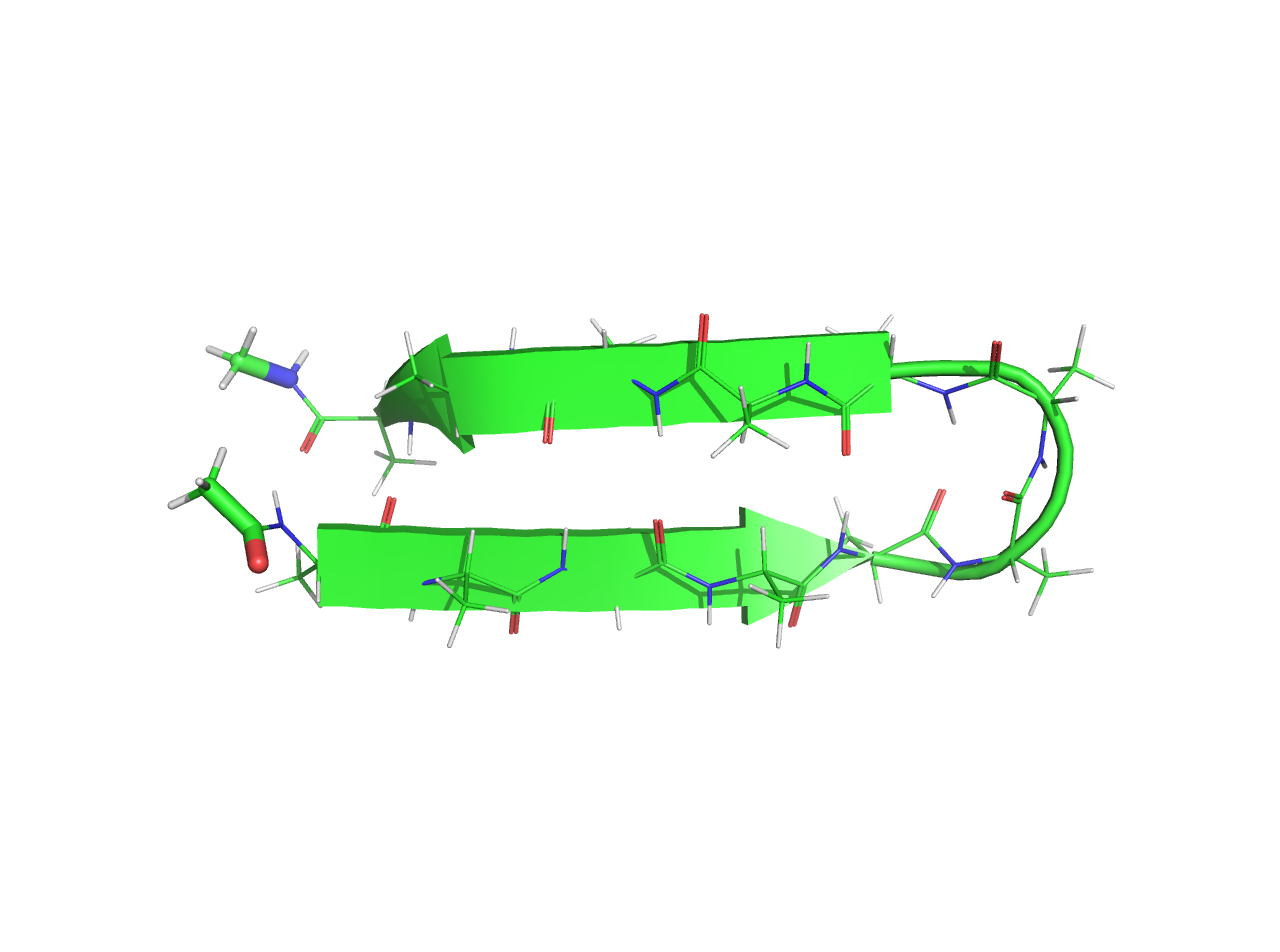}
                         \caption{B2}
         \end{subfigure}
          \caption{Folding of deca-alanine: three metastable conformations. } 
 \label{alanine12}
\end{figure}

As an illustration, consider the
problem of exploring all folded configurations of a deca-alanine
molecule in vacuum at a specific temperature (300K).  This short
peptide is a common model for illustrating the difficulties of folding
due its very complicated free energy
landscape~\cite{chipot2005exploring}.  Three typical states  are shown
in Figure \ref{alanine12}.  The many interactions between atoms of the
molecule mean that changes in structure are the result of a sequence
of coordinated moves. Because the transitions between the states occur
infrequently (they are `rare events') the results of short Langevin dynamics
simulation will typically constitute a highly localised exploration of
the region surrounding one particular conformational minimum.   

The
goal of enhanced sampling strategies is
to dramatically expand the range of observed states by modifying in some way the dynamical model.  Typical schemes \cite{torrie1977nonphysical,rosso2002use,maragliano2006temperature,ciccotti2005blue,laio2002escaping,comer2014adaptive,darve2001calculating} rely on the definition of collective variables in order to drive the enhanced sampling of the system only in relevant low dimensional coordinates describing the slowest time scales.

In this article, we discuss the automatic identification of collective variables for the purpose of enhancing sampling in applications such as molecular dynamics.   In some cases, the natural collective variables relate to underlying physical processes and can be chosen using scientific intuition, but in many cases, this is far from straightforward, as there may be competing molecular mechanisms underpinning a given conformational change.   Methods capable of automatically detecting CVs have, moreover, much wider potential for application.  Many Bayesian statistical inference calculations arising in clustering and classifying data sets, and in the training of artificial neural networks, reduce to sampling a smooth probability distribution in high dimension and are frequently treated using the techniques of statistical physics~\cite{murphy2013machine,chopin2012free}.  In such systems, a priori knowledge of the collective variables is typically not available, so methods that can automatically determine collective variables are of high potential value.

Diffusion maps~\cite{Coifman2006,coifman2008diffusion} provide a dimensionality reduction technique which yields a 
parameterized description of the underlying low-dimensional manifold by computing an approximation of a Fokker-Planck operator on the trajectory point-cloud sampled from a probability distribution (typically the Boltzmann-Gibbs distribution corresponding to prescribed temperature). The construction is based on a normalized graph Laplacian matrix. In an appropriate limit (discussed below), the matrix converges (point-wise) to the generator of overdamped Langevin dynamics. The spectral decomposition of the diffusion map matrix thus yields an approximation of the continuous spectral problem on the point-cloud~\cite{nadler_springer} and leads to natural collective variables. 
Since the first appearance of diffusion maps~\cite{Coifman2006},  several improvements have been proposed including local scaling~\cite{Rohrdanz2011a}, variable bandwidth kernels\cite{Berry2016} and target measure maps (TMDmap)~\cite{Banisch2018}.The latter scheme extends diffusion maps on point-clouds obtained from a surrogate distribution, ideally one that is easier to sample from. Based on the idea of importance sampling, it can be used on biased trajectories, and improves the accuracy and application of diffusion maps in high dimensions~\cite{Banisch2018}. Diffusion maps can be combined with the variational approach to conformation dynamics in order to improve the approximation of the eigenfunctions and provide eigenvalues that relate directly to physical relaxation timescales \cite{boninsegna2015investigating}.

Diffusion maps underpin a number of algorithms that have been designed to learn the collective variable adaptively and thus enhance the dynamics in the learned slowest dynamics~\cite{Zheng2013,Preto2014,chiavazzo2017intrinsic,chen2018molecular}. These methods are based on iterative procedures whereby diffusion maps are employed as a tool to gradually uncover the intrinsic geometry of the local states and drive the sampling toward unexplored domains of the state space, either through sequential restarting~\cite{Zheng2013} or pushing~\cite{chiavazzo2017intrinsic} the trajectory from the border of the point-cloud in the direction given by the reduced coordinates. In~\cite{m2017tica} time structure-based independent component analysis was performed iteratively to identify the slowest degrees of freedom and to use metadynamics~\cite{laio2002escaping} to directly sample them.  All these methods try to gather local information about the metastable states to drive global sampling, using ad hoc principles.  
In this paper, we provide a rigorous perspective on the construction of diffusion maps within a metastable state by formalizing the concept of a local equilibrium based on the \textit{quasi-stationary distribution}\cite{collet2012quasi}. Moreover, we provide the analytic form of the operator which is obtained when metastable trajectories are used in computing diffusion maps.

 Diffusion maps can also be used to compute committor
 functions~\cite{Lai2018}, which play a central role in
 transition path theory~\cite{vanden2006transition}. The committor is
 a function which provides dynamical information about the connection
 between two metastable states and can thus be used as a reaction
 coordinate (importance sampling function for biasing or splitting
 methods, for example). Committors, or the 'commitment probabilities', were first introduced as 'splitting probability for ion-pair recombination' by Onsager~\cite{onsager1938initial} and appear for example as the definition of $p_{\text{fold}}$, the probability of protein folding~\cite{du1998transition,pande1998pathways}.
Markov state models can in principle be used to compute committor probabilities~\cite{prinz2011efficient}, but high dimensionality makes grid-based methods intractable. The finite temperature string method~\cite{weinan2005finite} approximates the committor on a quasi-one-dimensional reaction tube, which is possible under the assumption that the transition paths lie in regions of small measure compared to the whole sampled state space. 
The advantage of diffusion maps is that the approximation of the Fokker-Planck operator holds on the whole space, and therefore we can compute the committor outside the reaction tube. 
Diffusion maps have already been used to approximate committor
functions in~\cite{Lai2018,thiede2018galerkin}. In~\cite{Lai2018}, in
order to improve the approximation quality, a new method based on a
point-cloud discretization for Fokker-Planck operators is introduced (however without any convergence result).  A more recent work~\cite{thiede2018galerkin} uses diffusion maps for committor computations.  
Finally, we mention that artificial neural networks were used to solve for the
committor in~\cite{khoo2018solving}, although the approach is much different than that considered here.  

The main conceptual novelty of this article lies in the insight on the
local versus global perspective provided by the quasi-stationary
distribution.   Thus, to be precise, compared to the work of \cite{Lai2018}, we clarify the procedure for going from the local perspective to designing an enhanced sampling method and we apply our methods to much more complicated systems (only toy models are treated in \cite{Lai2018}).   The second, more practical, contribution is the demonstration of the use of diffusion maps to identify the metastable states and to directly compute the committor function.   We consider, for example, the use of the diffusion map as a diagnostic tool for transition out of a  metastable state.  A third contribution lies in drafting an enhanced sampling algorithm based on quasi-stationary distribution and diffusion maps.  A careful implementation and application for small biomolecules shows the relevance and potential of this methodology for practical applications.   Algorithm 1 detailed in Section 5 will provide a starting point for further investigations.  The current article serves to bridge works in the mathematical and computational science literatures, thus helps to establish foundations for future rigorously based sampling frameworks.

This paper is organized as follows: in Section~\ref{sec:diffmaps}, we start with the mathematical description of overdamped Langevin dynamics and diffusion maps. In Section~\ref{section qsd}, we formalize the application of diffusion maps to a local state using the quasi-stationary distribution. We present several examples illustrating the theoretical findings. In Section~\ref{committors}, we define committor probabilities and the diffusion map-based algorithm to compute them. We also apply our methodology to compute collective variables, metastable states and committors for various molecules, including the alanine dipeptide and a deca alanine system.  In Section \ref{sec: accelerated sampling}, we show how the quasi-stationary distribution can be used to reveal transitions between molecular conformations.  Finally, we conclude by taking up the question of how the quasi-stationary distribution can be used as a tool for the enhanced sampling of large scale molecular models, paving the way for a full implementation of the described methodology in software framework.

\section{Langevin dynamics and diffusion maps \label{sec:diffmaps}}

We begin with the mathematical description of overdamped Langevin dynamics, which is used to generate samples from the Boltzmann distribution. By introducing the generator of the Langevin process, we make a connection to the diffusion maps which is formalized in the following section. We review the construction of the original diffusion maps and define the target measure diffusion map, which removes some of its limitations. 
 
\paragraph{Langevin dynamics and the Boltzmann distribution}
We denote the configuration of the system by $x \in \mathcal{D}$,
where, depending on the application, $\mathcal{D} = \mathbb{R}^d$,
$\mathcal{D}$ is a subset of $\mathbb{R}^d$  or $\mathcal{D} = (\mathbb{R}/\mathbb{Z})^d$ for systems with periodic boundary conditions. Overdamped Langevin dynamics  is defined by the stochastic differential equation
\begin{equation}
  \label{eq: Langevin}
    dx_t  =   -\n V(x_t) \, dt  + \sqrt{{2/\beta}} \, dW_t,
\end{equation}
where $W_t$ is a standard $d$-dimensional Wiener process, $\beta > 0$ is the inverse temperature and $V (x)$ is the potential energy driving the diffusion process.
The {\em Boltzmann-Gibbs} measure is invariant under this dynamics:
\begin{equation}
\label{eq: invariant measure}
\mu(dx)=Z^{-1}\mathrm{e}^{-\beta V(x)}\,dx, \qquad Z=\int_{\mathcal{D}}\mathrm{e}^{-\beta V(x)}\,dx.
\end{equation} 
The ergodicity property is characterized by the almost sure (a.s.) convergence of the trajectory average of a smooth observable $\vp$ to the average over the phase space  with respect to a probability measure, in this case $\mu$:
\begin{equation}
\lim_{t \rightarrow\infty}\widehat{\vp}_t=\E_{\mu}(\vp)\quad \text{a.s.}, \qquad \widehat{\vp}_t:=\frac{1}{t}\int_0^t\vp(x_s) \, ds.
\label{eq: ergodic averages}
\end{equation}
The infinitesimal generator of the Markov process $(x_t)_{t\geq 0}$, a solution of~\eqref{eq: Langevin},  is the differential operator 
\begin{equation}
\Lop_{\rm \beta} = - \n V\cdot \n + \beta^{-1} \de,
\label{eq: generator modified Langevin}
\end{equation}
defined for example on the set of $C^{\infty}$ functions with compact
support. The fact that $\Lop_{\rm \beta}$ is the infinitesimal
generator of  $(x_t)_{t\geq 0}$ means that (see for example~\cite{lelievre2010free}):
\[
 \forall x, \quad\frac{d}{dt}\Big[\mathbb{E}(\vp(x_t)\mid x_0=x)\Big]\bigl|_{t=0} = \Lop_{\beta} \vp(x), 
\]
where $A$ is $C^{\infty}$ compactly supported function and $x_0=x$ is
the initial condition at time $t=0$. Another way to make a link
between the differential operator~\eqref{eq: generator modified
  Langevin} and the stochastic differential equation~\eqref{eq:
  Langevin} is to consider the law of the process $x_t$. Let us denote
by $\psi(t, x)$ the density of $x_t$ at time $t$. Then $\psi$ is a solution of the Fokker-Planck equation 
\[
\du_t \psi = \mathcal{L}_{\rm \beta} ^*\psi, \quad \psi(0) = \psi_0,
\]
where $\psi_0$ is the density of $x_0$ and  $\mathcal{L}_{\rm \beta} ^*$ is the $L^2$ adjoint of $\Lop_\beta$. Under the assumption on the smoothness of the potential and the compactness of the domain $\mathcal{D}$, the solution can be written as 
\begin{equation}
\label{eq: sum}
\psi(t,x) = \sum_{k=0}^{\infty}c_k{\rm e}^{\lambda_k t}\phi_k(x),
\end{equation}
with eigenvalues $\{\lambda_k\}_{k=0}^{\infty}$ with $\lambda_0=0 >
\lambda_1 \geq \lambda_2\geq \ldots$ and eigenfunctions
$\{\phi_k\}_{k=0}^{\infty}$ of $\Lop^*_{\rm \beta} $. The
eigenfunctions are smooth functions and the sum~\eqref{eq: sum}
converges uniformly in $x$ for all times
$t>t_0>0$~\cite{coifman2008diffusion}. The ergodicity implies that
$\psi(t,x)\rightarrow c_0 \phi_0(x)$ as $t\rightarrow \infty$, and
therefore the first term $c_0 \phi_0$ in the sum~\eqref{eq: sum} is equal to $\mu$. The convergence rate is determined by the next dominant eigenfunctions and eigenvalues. A $k$-dimensional diffusion map at time $t$ is a lower dimensional representation of the system defined as a nonlinear mapping of the state space to the Euclidean space with coordinates given by the first $k$ eigenfunctions: 
\[
G_{k}(t, x):=({\rm e}^{\lambda_1 t}\phi_1(x), \ldots, {\rm e}^{\lambda_k t}\phi_k(x)  ).
\]

The diffusion distance is then the Euclidean distance between the diffusion map coordinates (DC). 

Finally, we would like to stress that although the eigenfunctions of the kernel matrix do approximate point-wise the spatial
eigenfunctions of the Fokker-Planck operator, there is no relationship between the
diffusion map matrix eigenvalues and the exact propagator eigenvalues as showed in~\cite{boninsegna2015investigating}. This connection would be provided by constructing diffusion map based approximator of the generator of underdamped Langevin dynamics (with finite friction).

\paragraph{Diffusion maps}

The diffusion map~\cite{Coifman2006} reveals the geometric structure of a manifold $\op{M}$ from given data~$\set{D}^{(m)}$ by constructing a $m\times m$ matrix that approximates a differential operator. The relevant geometric features of $\op{M}$ are expressed in terms of the dominant eigenfunctions of this operator.  

The construction requires that a set of points $\set{D}^{(m)} := \{x_1,x_2,\ldots,x_m\}\subset\R^N$ ($N>0$) which have been sampled from a distribution $\pi(x)$ lie on a compact $d$-dimensional differentiable submanifold $\mathcal{M}\subset \R^N$ with dimension $d<N$. 
The original diffusion maps introduced in \cite{Coifman2006,Singer2006} are based on the
isotropic kernel $h_{\e}(x, y) = h( \|x - y\|^2 / \e)$ where $h$ is an
exponentially decaying function, $\e > 0$ is a scale parameter and
$\|\cdot\|$ is a norm\footnote{For example Euclidean or RMSD, which is the most commonly used norm in molecular simualtions~\cite{boninsegna2015investigating}} in $\R^N$. A typical choice is 
\begin{equation}
h_{\e}(x,y) = \exp\left(-(4\e)^{-1}\|x-y\|^2\right).
\label{eq:kernel}
\end{equation}
In the next step, an $m\times m$ kernel matrix $K_{\e}$ is built by the evaluation of $h_{\e}$ on the set $\set{D}^{(m)}$. This matrix is then normalized several times to give a matrix $P_{\e}$ that can be interpreted as the generator of a Markov chain on the data. To be precise, the kernel matrix {$K_\e$}  is normalized using the power $\alpha\in [0,1]$ of the estimate $q$ of the density $\pi$, usually obtained from the kernel density estimate ${q_i}=\sum_{j=1}^NK_{ij}$ as the row sum of $K_{\e}$. In some cases the analytic expression of the density $\pi(x)$ is known and we can set directly $q(x) =\pi(x)$. After obtaining the transition matrix 
\begin{equation}
P_{\e} = D^{-1}_{\alpha}K_\e,
\label{P}
\end{equation}
where $D_{\alpha} = { {\rm diag}(q^{-\alpha})}$, we compute in the last step the normalized graph Laplacian matrix 
\begin{equation}
\label{graph laplacian}
L_{\e}=\e^{-1}(P_{\e}-I).
\end{equation}

As reviewed in~\cite{nadler_springer}, for given $\alpha$ and sufficiently smooth functions $f$, the matrix $L_{\e}$ converges in the limit $m\rightarrow \infty$ point-wise to an integral operator describing random walk in continuous space and discrete time, which in the limit $\e \rightarrow 0$ converges to the infinitesimal generator of the diffusion process in continuous space and time. Using the notation $[f] = (f(x_1),...,f(x_m))^T$ for representing functions evaluated on the data set  $\set{D}^{(m)}$ as vectors such that $[f]_i = f(x_i)$, we formally write the point-wise convergence: for $\alpha\in[0, 1]$, for $m\rightarrow \infty$ and $\e \rightarrow 0$, 
\[
(L_{\e} [f])_j \rightarrow \op{L}f(x_j), \hspace{0.1in} \text{ for all } x_j \in \set{D}^{(m)}, 
\] 
where $\Lop$ is a 
Fokker-Planck operator 
\begin{equation}
\label{eq:gen_diffmaps_vanilla}
\op{L}f = \Delta f + (2-2\alpha)\nabla f\cdot \frac{\nabla \pi}{\pi}
\end{equation}
where $\Delta$ is the Laplace-Beltrami operator on $\op{M}$ and
$\nabla$ is the gradient operator on $\op{M}$.  Note that in the special case $\alpha=1/2$, and for the choice $\pi =Z^{-1} {\rm e}^{-\beta V}$ (Boltzmann-Gibbs), the approximated operator corresponds to the generator of overdamped Langevin dynamics~\eqref{eq: generator modified Langevin}, such that
\begin{equation}
\label{eq:gen_langevin}
\op{L}=\beta \Lop_{\beta}.
\end{equation}
For this reason, we focus on the choice $\alpha=1/2$ throughout this work.

 Consequently, if there are enough data points for accurate statistical sampling, eigenvectors of $L_{\e}$ approximate discretized eigenfunctions of $\op{L}$. Then eigenvectors of $L_{\e}$ approximate solutions to the eigenproblem associated to $\op{L}$ 
 : $L_{\e}[\psi]=\lambda[\psi]$, an approximation of
 \[
 \Lop \psi(x)=\lambda \psi(x), \forall x\in \mathcal{M}\subset {\rm supp}(\pi).
 \]

The spectral decomposition of $L_{\e}$ provides real, non-positive eigenvalues $0 = \lambda_0 > \lambda_1 \geq \lambda_2 \geq \ldots \geq \lambda_m$ sorted in decreasing order. The dominant 
eigenfunctions allow for a structure preserving embedding $\Psi$ of $\set{D}^{(m)}$ into a lower dimensional space and hence reveal the geometry of the data.

Singer \cite{Singer2006} showed that for uniform density $\pi$, the approximation error for fixed $\e$ and $m$ is
\begin{equation}
(L_{\e} [f])_j = \Lop f(x_j)  + O\left(\e, m^{-1/2}\e^{-1/2-d/4} \right).
\label{error}
\end{equation}

\begin{remark}[Infinite friction limit]
	\label{remark inifinite friction limit} In molecular dynamics, trajectories are usually obtained by discretizing~\cite{leimkuhler2015computation} the (underdamped) Langevin dynamics:  
\begin{equation}
  \label{underdamped Langevin}
  \left\{
  \begin{aligned}
    dx_t & = M^{-1}p_t \, dt, \\
    dp_t & = -\n V(x_t) \, dt - \gamma M^{-1} p_t \, dt + \sqrt{\frac{2\gamma}{\beta}} \, dW_t,
  \end{aligned}
\right.
\end{equation}
where $x\in \R^d$ are positions, $p\in \R^d$ momenta, $W_t$ is a standard $d$-dimensional Wiener process, $\beta > 0$ is proportional to the inverse temperature, $M={\rm diag}{m_1, \ldots, m_d}$ is the diagonal matrix of masses, and $\gamma>0$ is the friction constant. The generator of this process is 
\[
\mathcal{L}_{\gamma} = M^{-1}p \cdot \n_x - \n V(x)\cdot \n_p +\gamma\left(-M^{-1}p\cdot \n_p + \frac{1}{\beta}\de_p\right).
\]
Recall that the diffusion maps approximate the
generator~\eqref{eq:gen_diffmaps_vanilla} which is the generator of
the overdamped Langevin dynamics, an infinite-friction-limit dynamics
of the Langevin dynamics~\eqref{underdamped Langevin}. Diffusion maps
therefore provide the dynamical information in  the large friction
limit $\gamma \rightarrow +\infty$, rescaling time as $\gamma t$,  or
in a small mass limit  ($M \to 0$).
\end{remark}
\begin{remark}
Note that diffusion maps require the data to be distributed with
respect to $\pi(x) \, dx$, which appears in the limiting
operator~\eqref{eq:gen_diffmaps_vanilla}. It implies that even though
diffusion maps eventually provide an approximation of the generator of
overdamped Langevin dynamics~\eqref{eq:gen_langevin}, one can use
trajectories from any ergodic discretization of underdamped
Langevin dynamics to approximate the configurational marginal
$\pi(x) \, dx$ as for example the BAOAB integrator \cite{Leimkuhler2013a}.
\end{remark}

\paragraph{The target measure diffusion map}
 If $\pi$ is particularly difficult to sample, it might be desirable to  use points which are not necessarily distributed with respect to $\pi$ in order to compute an approximation to the operator 
\begin{equation}
\label{eq: target density operator}
\mathcal{L} = \nabla \log(\pi) \cdot \nabla  +\Delta.
\end{equation}
For this purpose, the target measure diffusion map was recently introduced \cite{Banisch2018}. The main advantage is that it allows construction of an approximation of $\mathcal{L}$ even if the data points are distributed with respect to some distribution $\mu$ such that ${\rm supp }  \ \pi \subset {\rm supp }  \  \mu$. The main idea is to use the previously introduced kernel density estimator $q_{\e}$, already obtained as an average of the kernel matrix
$K_{\e}$. Since we know the density of the target distribution $\pi$, the matrix normalization step can be done by re-weighting $q$ with respect to the target density, which allows for the matrix normalization in an importance sampling sense.
 More precisely, the target measure diffusion map (TMDmap) is constructed as follows: 
The construction begins with the $m\times m$ kernel matrix $K_\e$ with components $(K_\e)_{ij} = k_\e(x_i, x_j)$ and the kernel density estimate $q_\e(x_i) = \sum_{j=1}^m (K_\e)_{ij}$.
 Then the diagonal matrix $D_{\e,\pi}$ with components $(D_{\e,\pi})_{ii} = \pi^{1/2}(x_i)q^{-1}_\e(x_i)$ is formed and the kernel matrix is right normalized with $D_{\e,\pi}$:
\[
K_{\e,\pi} = K_\e D_{\e,\pi}.
\]
Let us define $\tilde D_{\e,\pi}$ as the diagonal matrix of row sums of $K_{\e,\pi}$, that is,
\[
(\tilde D_{\e,\pi})_{ii} = (K_{\e,\pi} [\mathbf{1}])_i = \sum_{j=1}^m (K_{\e, \pi})_{ij}.
\]
Finally, the TMDmap matrix is built as
\begin{equation}
\label{eq:TMDmap}
L_{\e, \pi} = \e^{-1}\left( \tilde D_{\e, \pi}^{-1} K_{\e,\pi} - I\right).
\end{equation}
In~\cite{Banisch2018}, it is shown that $L_{\e, \pi} $ converges
point-wise to the operator~\eqref{eq: target density operator}.

\begin{remark}[Nystr\"om extension of the eigenvectors \cite{bengio2004learning}] Note that the $j$-th eigenvector $[\psi]$ of the matrix $L_{\e}$ can be extended on $x\in \mathcal{M}$ as
\[
\psi_j(x) =\frac{1}{\lambda_j}\sum_{i=1}^N\frac{h_{\e}(x, x_i)}{D(x)}[\psi_j]_i
\]
where $\lambda_j\neq 0$ is the corresponding eigenvalue, $D(x) = \sum_{i=1}^N h_{\e}(x, x_i)$ and $[\psi_j]_i = \psi_j(x_i)$.
\end{remark}

\paragraph{Dirichlet boundary problems}

Diffusion maps provide a matrix $L_{\e}$, which converges point-wise
to the generator $\Lop$ defined
in~\eqref{eq:gen_diffmaps_vanilla}. This method can be used to solve
the following eigenvalue problem with homogeneous Dirichlet boundary conditions: find $(\lambda, f)$ such that $\Lop f = \lambda f  \text{ in }\Omega, f=0 \text{ in }\du \Omega.$
In the following example, we solve a linear eigenvalue problem with Dirichlet boundary conditions.  In the first step, we construct $L_{\e}$ as the diffusion map approximation of $\Lop$ on the point-cloud $\{x_i\}_{i=1}^m$. In order to express the Dirichlet boundary condition, we identify points outside the domain $\Omega$ that we define as $\mathcal{C}:=\{x_j\}_{j\in J}$ where the set of indices $J:= \{j:  x_j \notin \Omega\}$. Finally,  we solve the eigenvalue problem with matrix $L_{\e}$, in which rows  with indices in $J$ have been set to zero. 

\begin{figure}
    \centering
    \begin{subfigure}[b]{0.45\textwidth}
        \includegraphics[width=\textwidth]{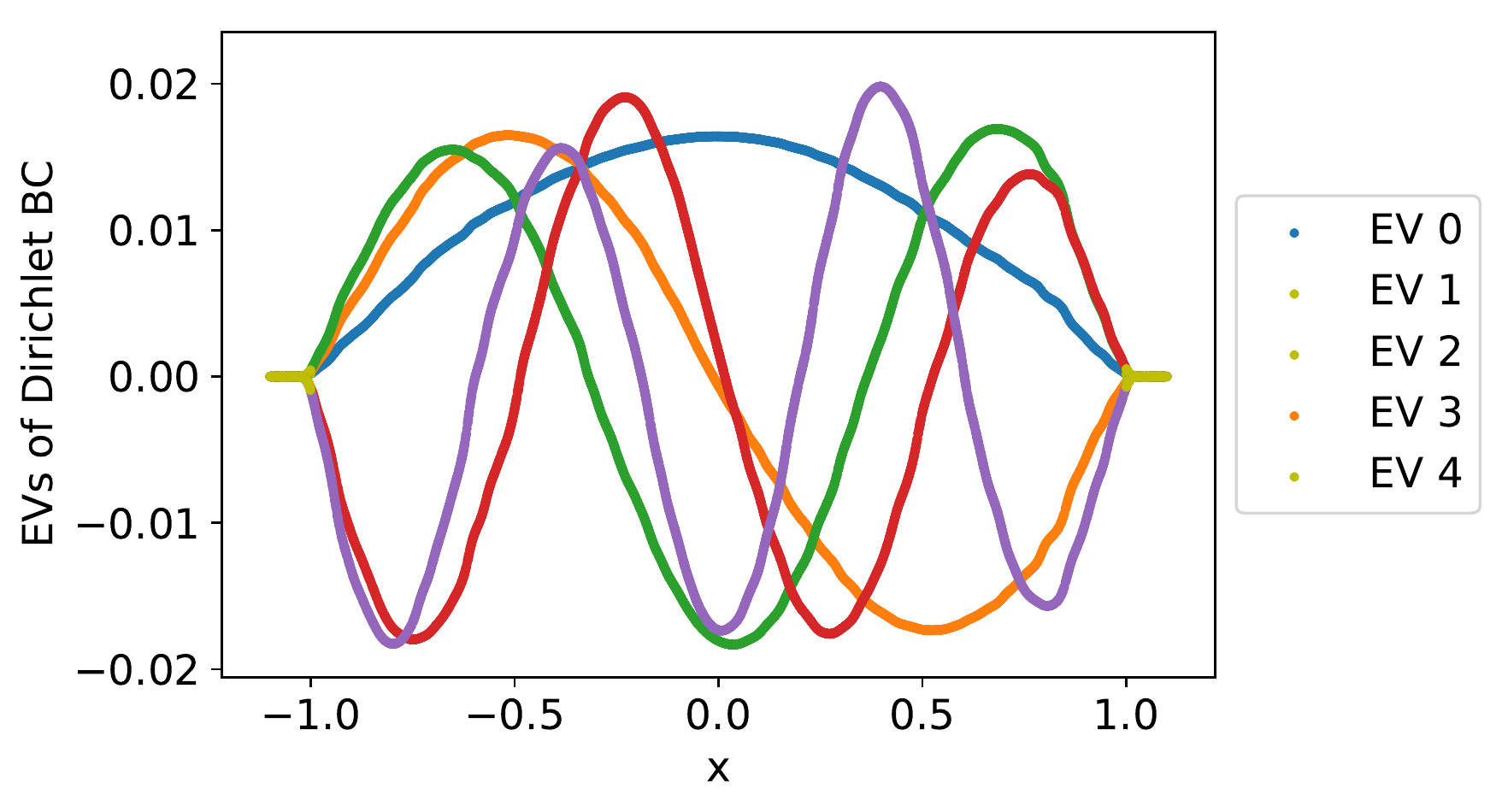}
    \end{subfigure}
   ~
    \begin{subfigure}[b]{0.45\textwidth}
        \includegraphics[width=\textwidth]{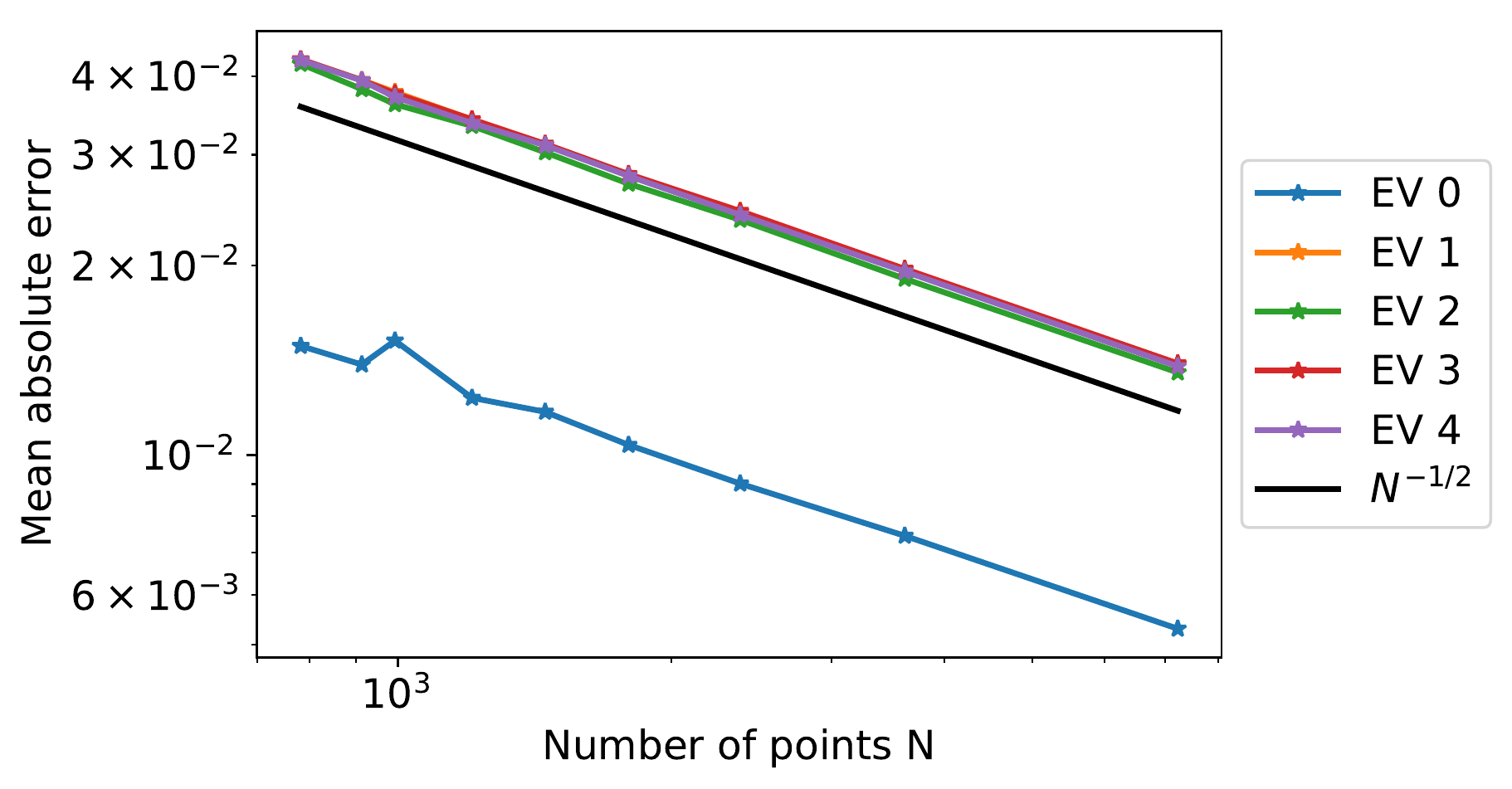}
    \end{subfigure}    
 \caption{(a) Eigenvectors obtained from diffusion maps. (b) Mean absolute error on the first five eigenfunctions over the number of samples. }
 \label{convergence_gaussian_eigenvectors}      
\end{figure}

Let us illustrate this on the following one-dimensional eigenvalue problem
\begin{equation}
\Lop v= \lambda v, \quad v\in (-1, 1), \quad v  = 0, v \in \{-1, 1\},
\label{dirichlet1d}
\end{equation}
with $ \Lop = - x \du  +\du^2 $. The eigenfunctions are $\psi_k(x) =
\frac 1 2 \sin(\pi(x+1)(k+1)$. To approximate the solution of~\eqref{dirichlet1d}  using diffusion maps with $\alpha=1/2$, we have generated $10^6$ points from a discretized overdamped Langevin trajectory with potential $V(x) = x^2/2$, using a second order numerical scheme~\cite{Leimkuhler2013a}. In Figure~\ref{convergence_gaussian_eigenvectors}(a) we show the diffusion map approximation of the eigenfunctions $\psi_k/\| \psi_k\|_2$.  From Figure~\ref{convergence_gaussian_eigenvectors}(right) we observe that the decay  of the mean absolute error of the normalized eigenfunctions is asymptotically proportional to ${N}^{-1/2}$, where $N$ is the number of samples. Different numbers $N$ of samples were obtained by sub-sampling the trajectory.

\section{Defining a 'local' perspective in diffusion-map analysis}
\label{section qsd}

We now concentrate on the case when diffusion maps are built using
trajectories of the Langevin dynamics. As we have reviewed in the
introduction, many iterative methods aim at gradually uncovering the
intrinsic geometry of the local states. The information obtained from
the metastable states can then be used, for example, to accelerate the sampling towards unexplored domains of the state space. 

The approximation error of  diffusion maps~\eqref{error} scales in
$O(m^{-1/2})$, $m$ being the number of samples. In order to provide an
approximation of a Fokker-Planck operator~\eqref{eq: generator
  modified Langevin} using a point-cloud obtained from the process
$x_t$, a solution of (a discretized version of)~\eqref{eq: Langevin},
the time averages should have converged with a sufficiently small
statistical error. In this section, using the notion of the
quasi-stationary distribution, we explain to which operator converges
a diffusion map approximation constructed on the samples in a metastable subset of the state space and why it is possible to obtain convergence of the approximation in this setup. 

\paragraph{Quasi-stationary distribution}
The quasi-stationary distribution (QSD) is a local equilibrium macro-state describing a process trapped in a metastable state $\Omega \subset  \R^{d}$ (see for example~\cite{collet2012quasi}). The QSD $\nu$ can be defined as follows: for all smooth test functions $\vp: \R^{d}\rightarrow \R$,
\[
\forall X_0 \in \Omega, \, \lim_{t\rightarrow \infty} \E \left( \vp \left(X_t)\right)\mid \tau > t \right)  = \int_{\Omega}\vp d\nu, 
\]
where $\Omega$ is a smooth bounded domain in $\R^{d}$, which is the
support of $\nu$ and the first exit time $\tau$ from $\Omega$ for $X_t$ is defined by
\[
\tau =\inf \{t>0: X_t \not\in \Omega \}.
\]
The following mathematical properties of the QSD were proved in~\cite{le2012mathematical}.
The probability distribution $\nu$ has a density $v$ with respect to
the Boltzmann-Gibbs measure $\pi(dx) = Z^{-1} {\rm e }^{-\beta V(x)} dx$. The density $v$ is the first eigenfunction  of the infinitesimal generator $\Lop$ of the process $X_t$, with Dirichlet boundary conditions on $\du \Omega$:
\begin{equation}
\left\{
\begin{aligned}
\Lop v &=-\lambda v, & {\rm in} \  \Omega, \\
v &=0, & {\rm on} \  \du\Omega,
\end{aligned}
\right.
\label{ev pb qsd}
\end{equation}
where $-\lambda < 0$ is the first eigenvalue. The density of $\nu$
with respect to the Lebesgue measure is thus
\begin{equation}
 \forall x\in\Omega, \quad \nu(x) = \frac{v(x) {\rm e}^{-\beta V(x)}}{\int_{\Omega}v(x) {\rm e}^{-\beta V(x)}dx}.
 \label{qsd}
\end{equation}

Let us consider the situation when $\Omega$ is a metastable state for
the dynamics $(X_t)$, and $X_0 \in \Omega$. Then, for a long time,
the process remains in $\Omega$. If the diffusion map is built using
those samples in $\Omega$, it then provides an approximation of the
Kolmogorov operator~\eqref{eq:gen_diffmaps_vanilla} where $\pi$ is
replaced by the quasi-stationary distribution $\nu$, namely
\begin{equation}
\op{L}_{\Omega}f := \Delta f + (2-2\alpha)\nabla f\cdot \frac{\nabla \nu}{\nu} =  \Delta f + (2-2\alpha)\nabla f\cdot \left( \nabla {\rm ln}(v) - \beta \nabla V\right), \quad \text{on supp}(\Omega).
\label{eq: qsd operator}
\end{equation}
Notice that if $\Omega = \mathcal D$, $\nu=\pi$ and  we recover the operator~\eqref{eq:gen_diffmaps_vanilla} with respect to the distribution  $\pi$. 
In the case when $\Omega$ is in the basin of attraction of a local
minimum $x_0$ of $V$ for the dynamics $\dot{x} = -\nabla V (x)$, the
quasi-stationary distribution with density $\nu$ defined
by~\eqref{qsd} is exponentially close to $\pi(x) = Z^{-1}{\rm
  e}^{-\beta V(x)}dx$ on any compact in $\Omega$: the two
distributions differ essentially on the boundary $\partial \Omega$.  More precisely, as proved in~\cite[Lemma 23, Lemma 85]{di2017sharp} for example (see also~\cite[Theorem 3.2.3]{HELFFER_NIER}, for any compact subset $K$ of $\Omega$, there exists $c > 0$ such that, in the limit $\beta \to \infty$,
\[ \left\| \frac{1_K \exp(-\beta V)}{\int_K \exp(-\beta V)} - \nu \right\|_{L^2(\Omega)} = {\mathcal O}( {\rm exp}(-\beta c)).\]

In order to illustrate these ideas, we compare the diffusion map
constructed from a trajectory in a metastable state and points from a
trajectory which has covered the whole support of the underlying
distribution. As explained above, the distribution of the samples in
the metastable state is the quasi-stationary distribution and
diffusion maps provide an approximation of the operator~\eqref{eq: qsd operator}.

\begin{figure}
    \centering
    
        \includegraphics[width=\textwidth]{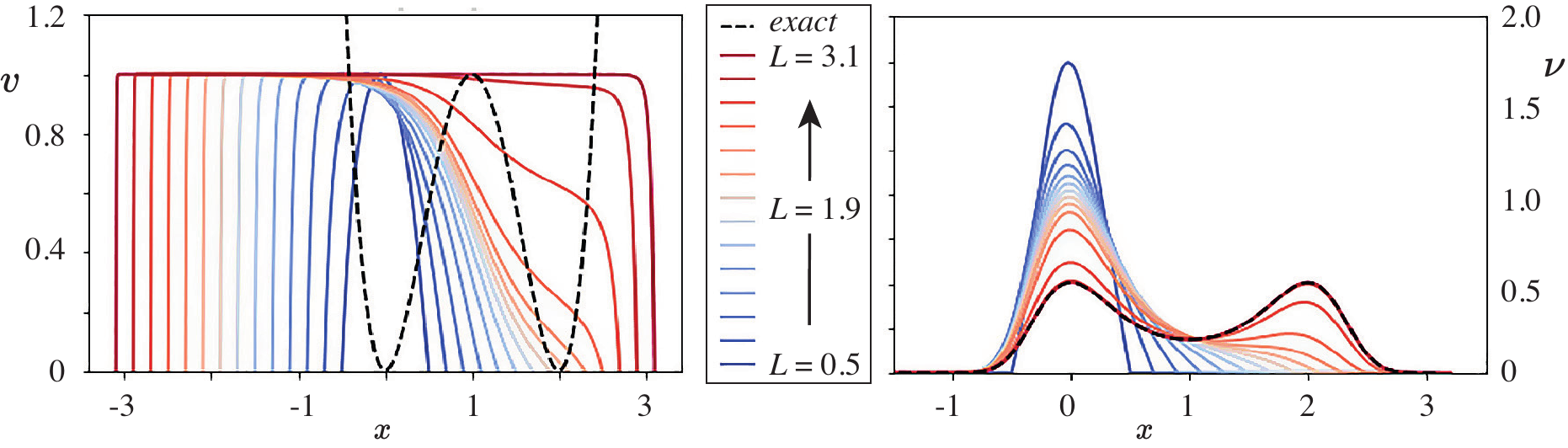}

 \caption{Quadratic potential in 1D.  (left) The approximation of the first eigenfunction $v$ of Dirichlet boundary eigenvalue problem~\eqref{ev pb qsd}. (right) The quasi-stationary-distribution $\nu$.}
 \label{fig:distr_qsd}      
\end{figure}

Let us next illustrate the quasi-stationary distribution and explain
how it differs from the stationary distribution on a simple one-dimensional example. The density of the quasi-stationary distribution~\eqref{qsd} can be obtained using accurate numerical approximation of the solution $v$ of the Dirichlet problem~\eqref{ev pb qsd} by a pseudo-spectral Chebyschev method~\cite{trefethen2000spectral,nold2017pseudospectral} in interval $[-L, L]$ with $L>0$, with the grid chosen fine enough to provide sufficient numerical accuracy. 
We consider a simple double-well potential $V(x) = ((x-1)^2-1)^2$. 
To illustrate the convergence of the quasi-stationary distribution $\nu = Z^{-1}_{\nu}v{\rm e}^{-V}$ to $\pi = Z^{-1}{\rm e}^{-V}$ we increase the interval size by plotting the approximation for increasing values of $L$. In  Figure~\ref{fig:distr_qsd} (a), we plot the approximation of the solution $v$ of the Dirichlet eigenvalue problem~\eqref{eq: qsd operator} on $[-L, L]$ for several values of $L>0$. As expected, we observe that $v(x)\rightarrow 1$ as $L\rightarrow \infty$. In Figure~\ref{fig:distr_qsd} (b), we plot the corresponding quasi-stationary densities. As the size of the domain increases, the quasi-stationary distribution converges to $\pi$. 



\begin{figure}
    \centering
    \begin{subfigure}[b]{0.45\textwidth}
        \includegraphics[width=\textwidth]{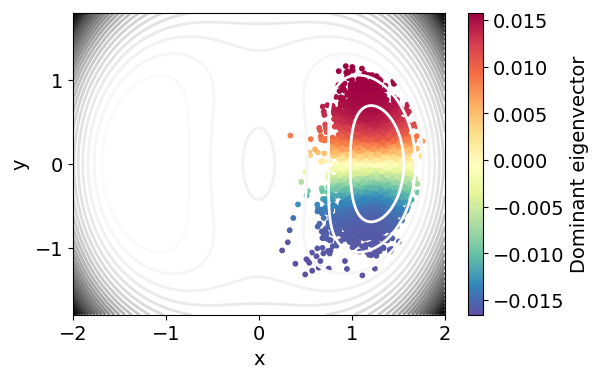}
    \end{subfigure}
   ~
    \begin{subfigure}[b]{0.45\textwidth}
        \includegraphics[width=\textwidth]{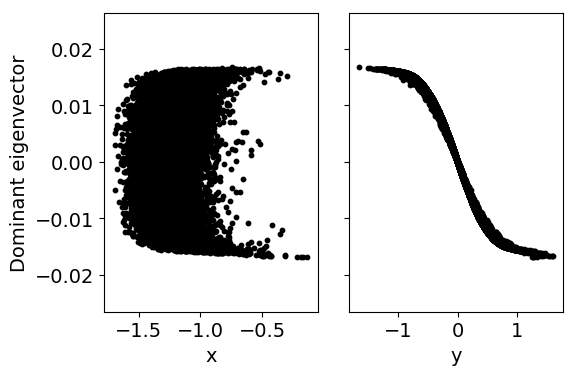}
    \end{subfigure}    
 \caption{Local geometry.  (a) The sampling of the metastable state.  (b) The dominant eigenvector parameterizes the $y$-coordinate.}
 \label{fig:DW2dcv_both_ev_0}      
\end{figure}

\begin{figure}
    \centering
    \begin{subfigure}[b]{0.45\textwidth}
        \includegraphics[width=\textwidth]{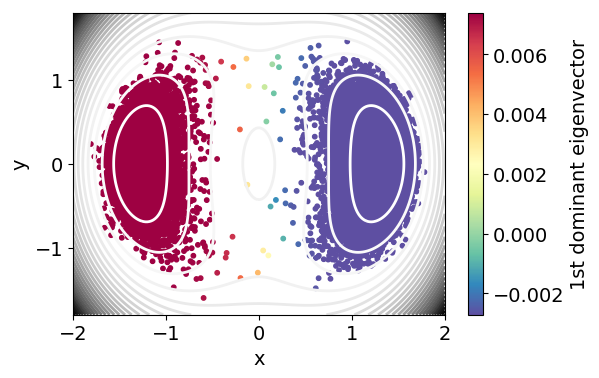}
    \end{subfigure}
   ~
    \begin{subfigure}[b]{0.45\textwidth}
        \includegraphics[width=\textwidth]{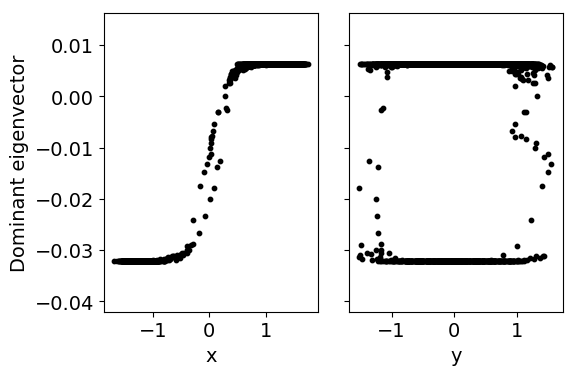}
    \end{subfigure}    
 \caption{Global geometry.  (a) The samples cover the whole support of the distribution.  (b) The dominant eigenvector correlates with the $x$-coordinate.}
  \label{fig:DW2dcv_both_ev_2}      
\end{figure}

In the next example, we sample from the Boltzmann distribution with a
two-dimensional double-well~\cite[Section
1.3.3.1]{lelievre2010free}:
\begin{equation}
V_{\rm DW} (x, y)= \frac{1}{6} (4 (-x^2-y^2+w)^2+2 h (x^2-2)^2+((x+y)^2-w)^2+((x-y)^2-w)^2),
\label{eq:DW}
\end{equation}
with $h=2$ and $w=1$. We employ a second-order discretization of
Langevin dynamics~\eqref{eq: Langevin} at low temperature ($\beta=10$).  Due to this low temperature, the
samples are trapped in the first well long enough to locally
equilibrate to the quasi-stationary distribution. We compute the
statistical error of averages of various observables such as the configurational and the kinetic temperatures, because of the knowledge of the exact expected values of these observables\footnote{The following equalities hold, respectively, for the kinetic and configurational temperatures: $k_BT = \E_{\pi}[p\cdot \nabla U(p)] = \frac{\E_{\pi}[|\nabla V(x)|^{2}]}{\E[\Delta V(x)]}$, where $k_B$ is the Boltzmann constant, $T$ is temperature  and $U(p)=\frac{1}{2}p^\intercal M^{-1}p$ is kinetic temperature.}. We also track the first dominant eigenvalues from the diffusion map, in order to detect when the sampling has reached a local equilibrium. 
From this short trajectory, the diffusion map uncovers the geometry of
the local state: the dominant eigenvector clearly parameterizes the
$y$-coordinate, i.e. the slowest coordinate within the metastable state, see Figure~\ref{fig:DW2dcv_both_ev_0}. On the other hand, when the trajectory explores also the second well and hence covers the support of the distribution, the diffusion map parameterizes $x$ as the slowest coordinate, see Figure~\ref{fig:DW2dcv_both_ev_2}.

These examples demonstrate that diffusion maps can be constructed using points of the quasistationary distribution to uncover the slowest local modes. As we will show in Section \ref{sec: accelerated sampling} this property will eventually allow us to define local collective variables which can guide the construction of sampling paths that exit the metastable state. 

\section{Global perspective: Identification of metastable states and committors}
\label{committors}

We illustrate in this section how diffusion maps applied to global sampling can be used to approximate the
generator of Langevin dynamics and
committor probabilities between metastable states in infinite-friction limit\footnote{We are aware that diffusion maps do not provide access to dynamical properties. However, having access to better reaction coordinate such as the committor can be used to obtain dynamical properties (for example in combination with Adaptive Multilevel Splitting\cite{cerou2007adaptive}).}.
  We also use diffusion maps to automatically identify metastable sets in high-dimensional systems.

The committor function is the central object of transition path
theory~\cite{vanden2006transition,metzner2006illustration}. It
provides a dynamical reaction coordinate between two metastable states $A\subset \mathcal{D}$ and $B\subset\mathcal{D}$. The committor function is the probability that the trajectory starting from $x\not\in A\cup B$  reaches first $B$ rather than $A$,
\[
q(x) = \mathbb{P}(\tau_B<\tau_A \mid x_0=x)
\]
 The committor $q$ is also the solution of the following system of equations:
\begin{equation}
\left\{
\begin{aligned}
\Lop q &=0,  {\rm in} \  \mathcal{D} \setminus (A \cup B), \\
q &=0, {\rm in} \  A, \\
q &=1,  {\rm in} \  B. 
\end{aligned}
\right.
\label{committor}
\end{equation}
From the committor function, one can compute the reaction rates, density and current of transition paths~\cite{lu2015reactive}.

In the spirit of~\cite{Lai2018}, we use diffusion maps to compute committors from the trajectory data. Given a point-cloud on $\mathcal{D}$, diffusion maps provide an approximation of the operator $\Lop$ and can be used to compute $q$.
After computing the graph Laplacian matrix~\eqref{graph laplacian}
(choosing again $\alpha=1/2$), we solve the following linear system,
which is a discretization of~\eqref{committor}:
\begin{equation}
L_{\e}[c,c] q[c]= -L_{\e}[c,b]q[b],
\label{committor linear system}
\end{equation}
where we defined by $c$ and $b$ indices of points belonging to the set $ C = \mathcal{D} \setminus (A \cup B)$ and $B$ respectively, and $L_{\e}[I, J]$ the projection of the matrix or vector on the set of indices $I=\{I_k\}_{k=1}^{\#I}$ and $J=\{J_k\}_{k=1}^{\#J}$.




\paragraph{One-dimensional Double-well.}

\begin{figure}
    \centering
    \begin{subfigure}[b]{0.45\textwidth}
        \includegraphics[width=\textwidth]{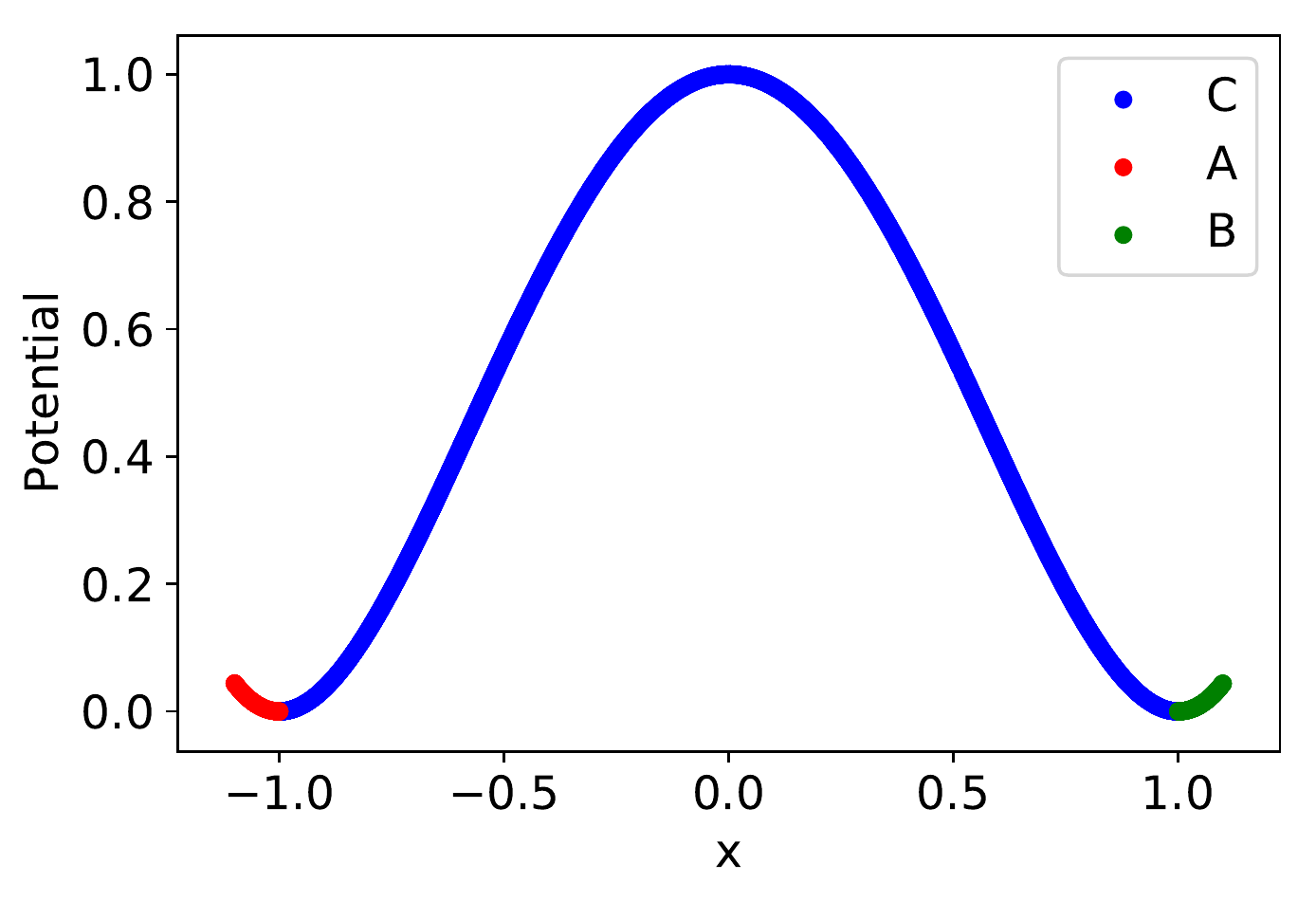}
    \end{subfigure}
   ~
    \begin{subfigure}[b]{0.45\textwidth}
        \includegraphics[width=\textwidth]{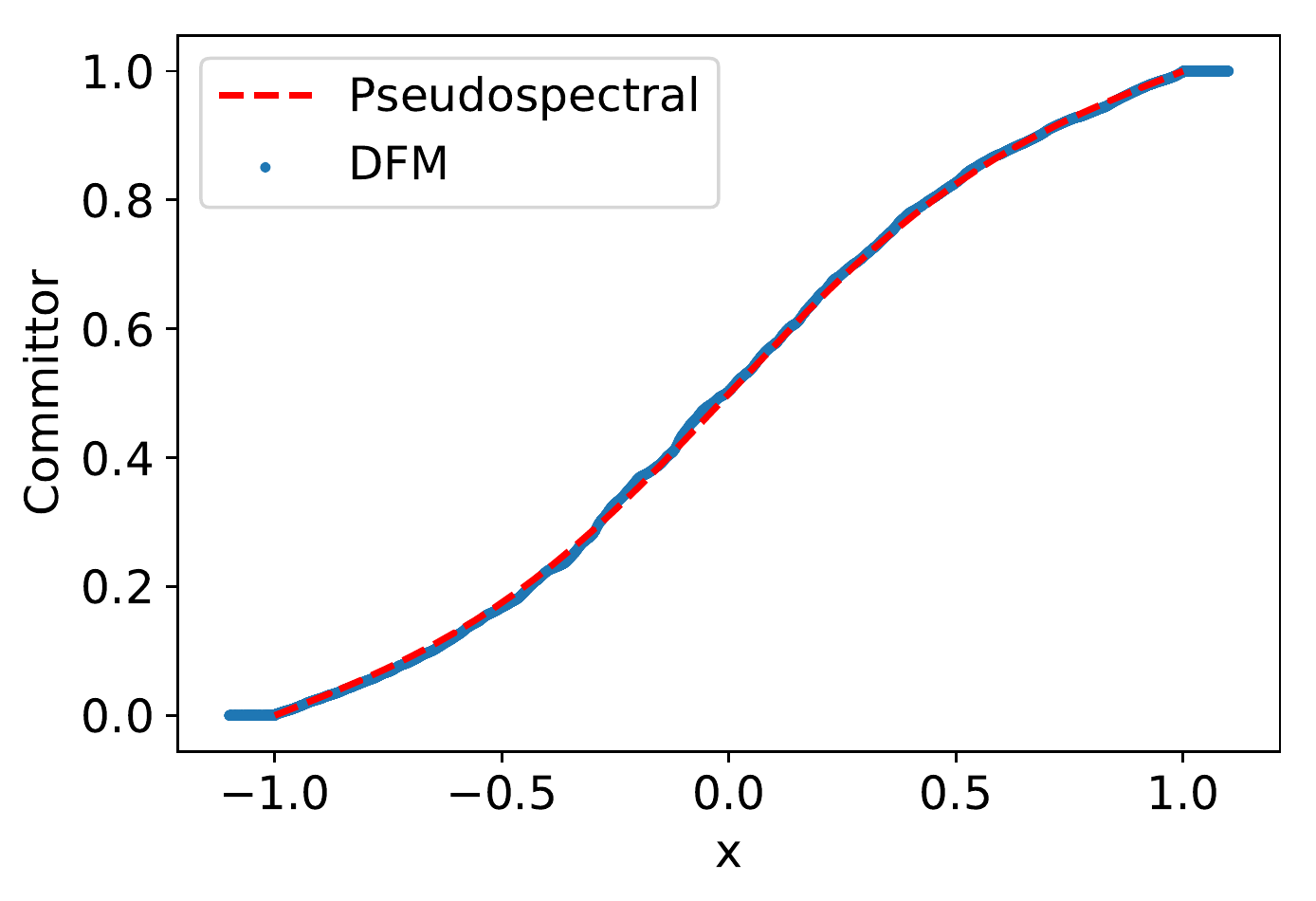}
    \end{subfigure}
 \caption{Double-well in 1D. (a) A plot of the potential showing the three sets $A, B$ and $C=\mathcal{D} \setminus (A \cup B)$. (b) The committor function approximation by diffusion maps (DFM) and an accurate pseudospectral method.}
 \label{gaussian_committor_1d}
\end{figure}

First, we compute the committor function for a one-dimensional double-well potential $V(x)=(x^2-1)^2$. We use a second order discretized Langevin scheme with step size $\de t=0.1$ to generate $10^5$ points and compute the TMDmap with $\epsilon=0.1$. We fix the sets $A=[-1.1,  -1]$ and $B=[1, 1.1]$, see   Figure~\ref{gaussian_committor_1d} (a). In Figure~\ref{gaussian_committor_1d} (b),  we  compare the committor approximation with a solution obtained with a pseudo-spectral method \cite{trefethen2000spectral}, which would be computationally too expensive in high dimensions.

\subsection{Algorithmic identification of metastable subsets}

The most commonly used method for identifying metastable subsests is  the Perron-cluster cluster analysis (PCCA), which exploits the structure of the eigenvectors \cite{schutte1999direct,schutte2003,prinz2011probing,deuflhard2005robust}.
In this work, we use the eigenvectors provided by diffusion maps in order to automatically identify the metastable subsets. 
The main idea is to compute the dominant eigenvectors of the transfer operator $P$, which are those with eigenvalues close to $1$, excluding the first eigenvalue. 
We approximate $P$ by $P_{\e}$ defined in~\eqref{P}. The metastable states can be clustered according to the 'sign' structure of the first dominant eigenvector, which is moreover constant on these states. More precisely, we find the maximal and the minimal points of the first eigenvector, which define the centers of the two sets. In the next step, we `grow' the sets by including points with Euclidean distance in diffusion space smaller than a fixed threshold.  See for example Figure~\ref{fig:DW2dcv_both_ev_2}(b) for an illustration in the case of the two dimensional double-well potential.

\paragraph{Two-dimensional Double-well}

\begin{figure}

    \centering
    \begin{subfigure}[b]{0.45\textwidth}
        \includegraphics[width=\textwidth]{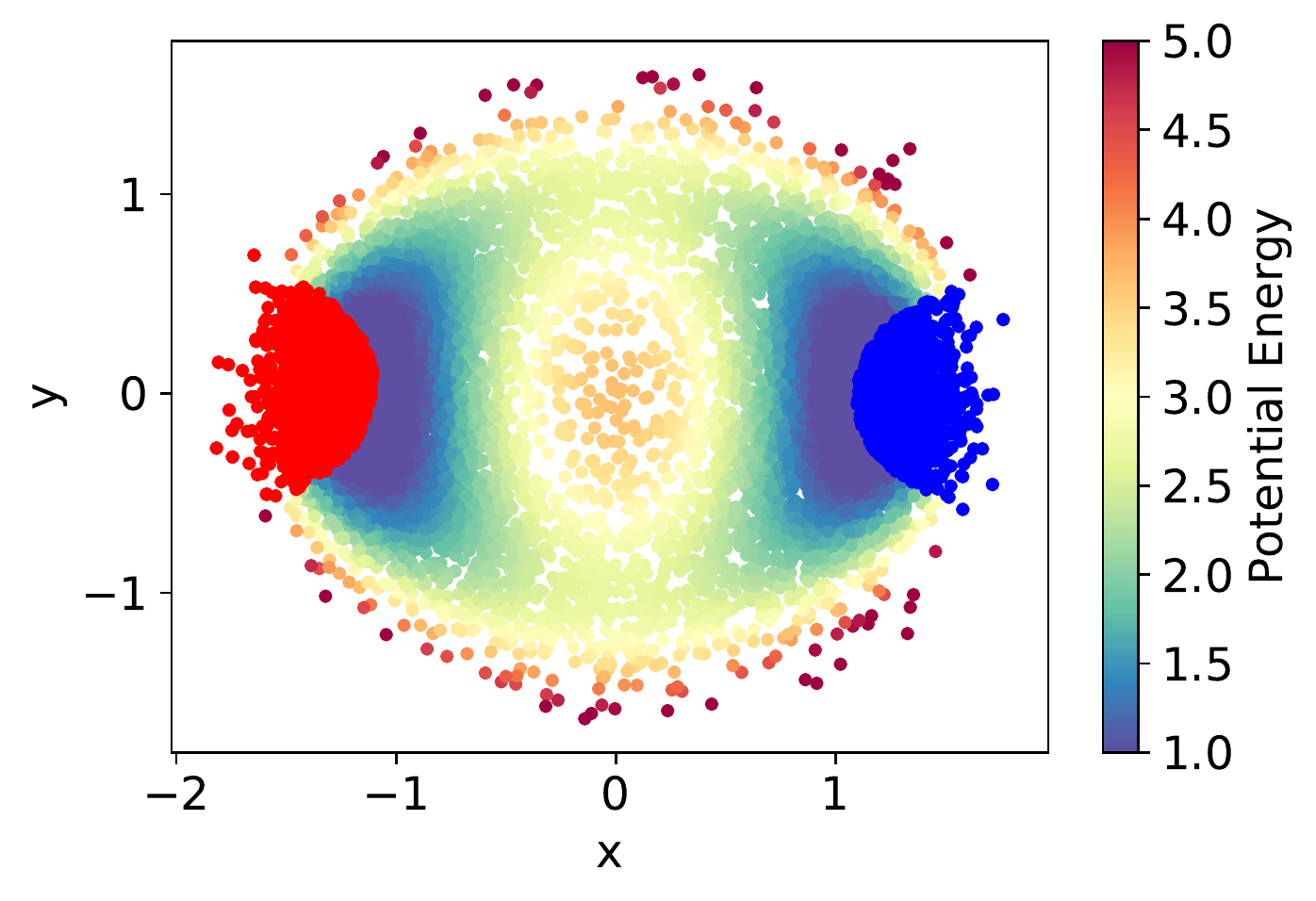}
        
    \end{subfigure}
   ~
    \begin{subfigure}[b]{0.45\textwidth}
        \includegraphics[width=\textwidth]{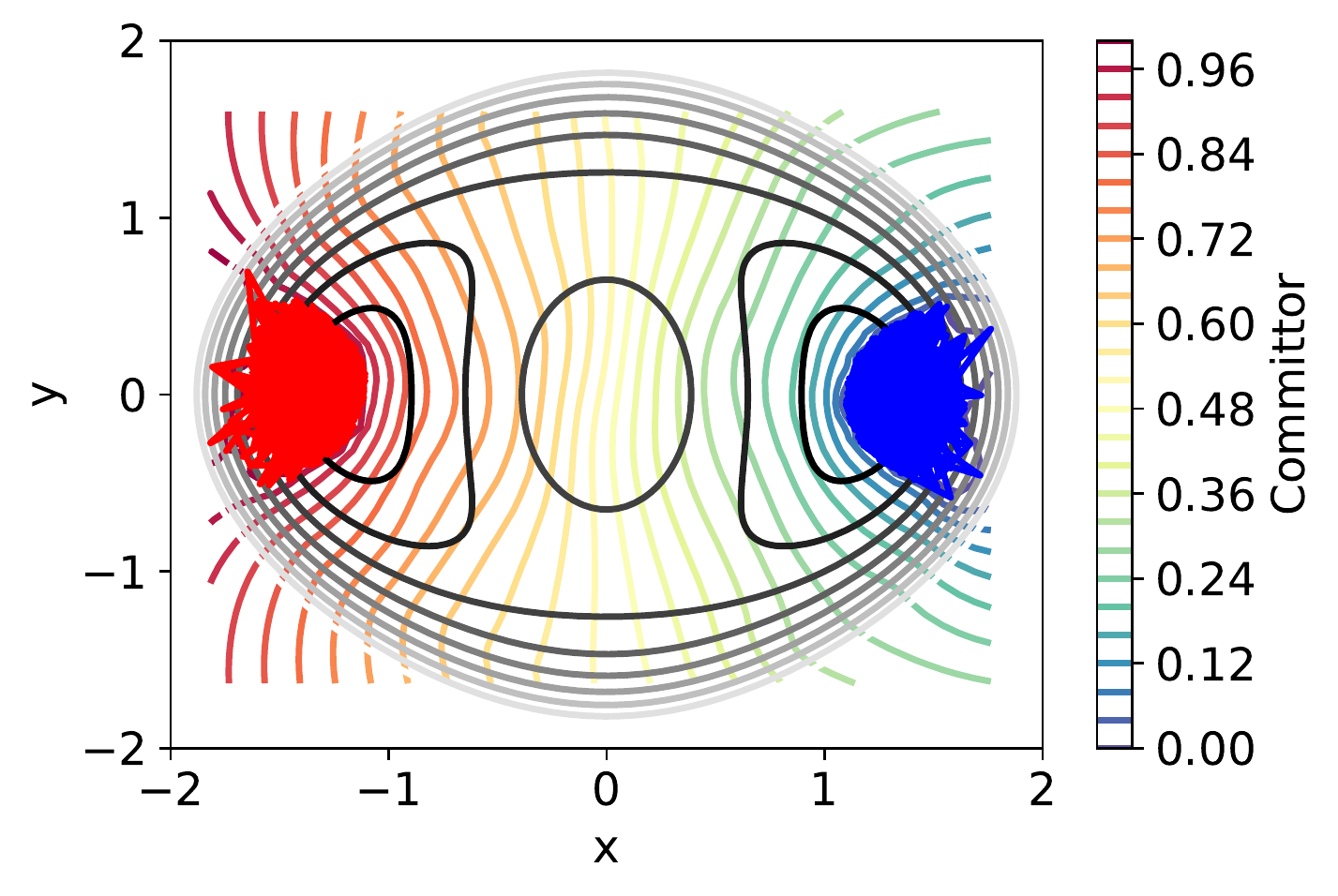}
    \end{subfigure}
 \caption{Two dimensional double-well potential: committor approximations on the automatically identified metastable sets. (a) The sampled trajectory with the automatically chosen sets set $A$ (blue) and $B$ (red). (b) The committor extended on a grid. Note that the value of $0.5$ is close to $x=0$ as expected.}
  \label{DW2d_proba_flow_automatic}
\end{figure}

\begin{figure}
    \begin{subfigure}[b]{0.45\textwidth}
        \includegraphics[width=\textwidth]{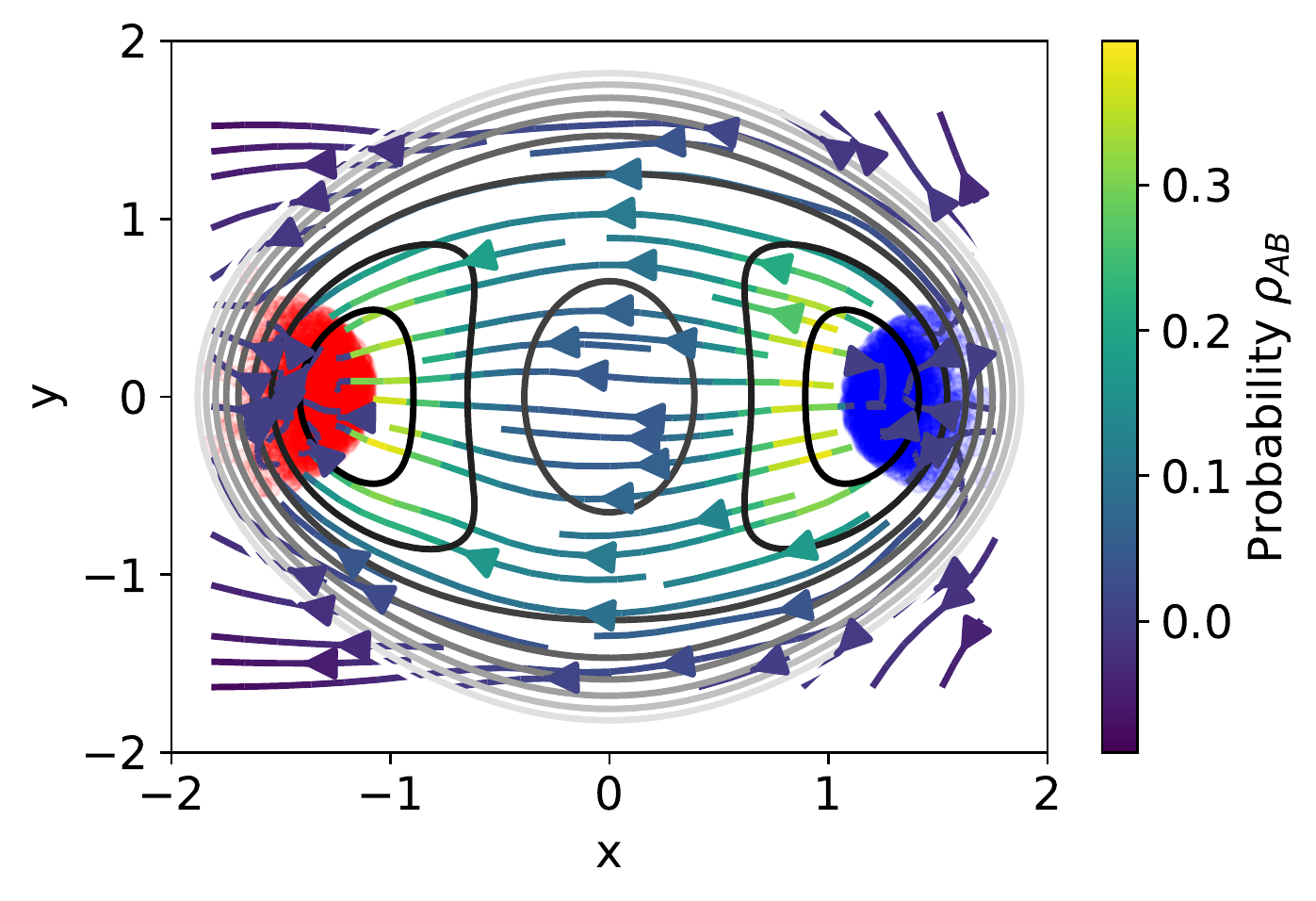}
    \end{subfigure}
     ~
 \begin{subfigure}[b]{0.45\textwidth}
        \includegraphics[width=\textwidth]{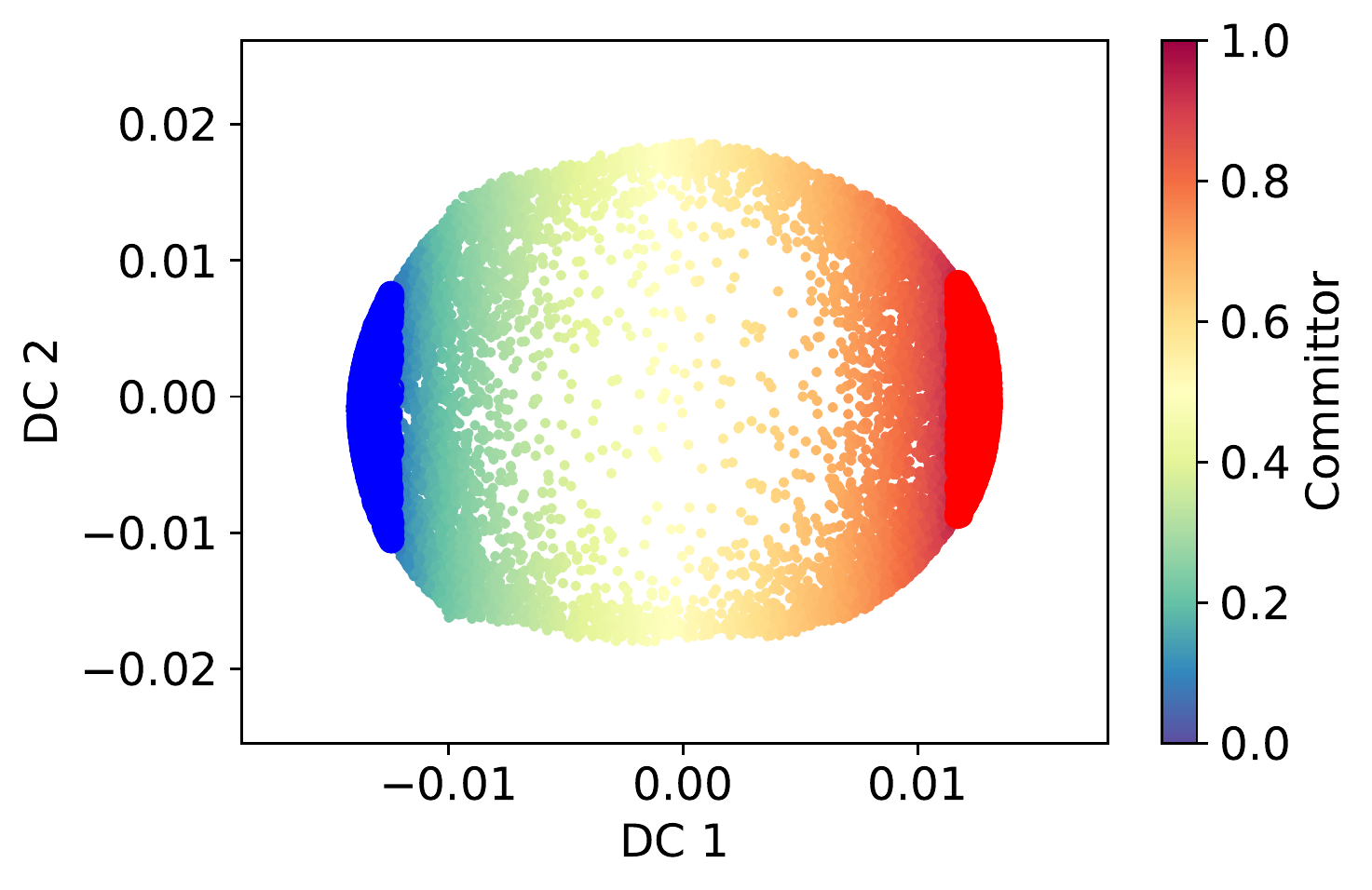}
                
    \end{subfigure} 

 \caption{(a) The flux streamlines coloured by probability~$\rho_{AB}$. (b) The diffusion map embedding and the chosen sets $A, B$ (colored blue and red).}
 \label{DW2d_proba_flow_automatic_2}
\end{figure}

In the second example, we consider a two-dimensional problem with potential~\eqref{eq:DW} with $h=2$ and $w=1$. We generate $m= 10^5$ samples using discretized Langevin dynamics with timestep $\de t=0.1$. We use $10^4$ points obtained by sub-sampling of the trajectory for the diffusion maps, which is chosen with kernel~\eqref{eq:kernel} and $\e=0.1$. We compute the first two dominant eigenvectors and define the metastable subsets $A$ and $B$ using the first dominant eigenvector as described above\footnote{ The results suggest that the method is robust with respect to the variation of the definition domains, under the assumption that the domains $A$ and $B$ are within the metastable state.}. In Figure~\ref{DW2d_proba_flow_automatic}(a) we see the sampling and the chosen sets. We compute the committor by solving the linear system~\eqref{committor linear system} and extrapolate linearly the solution on a two dimensional grid of $x$ and $y$; this is illustrated in the right panel of Figure~\ref{DW2d_proba_flow_automatic} (b). 
The representation of the committor in the diffusion coordinates is depicted in Figure~\ref{DW2d_proba_flow_automatic_2} (b). 

\subsection{High-dimensional systems}
\label{sec:molecules}

We now use diffusion maps to compute dominant eigenfunctions of the transition operator, identify metastable sets and approximate the committors of small molecules: alanine dipeptide and deca-alanine. We discuss the local and global perspective by computing diffusion maps inside the metastable states. For the example of deca-alanine, we use a trajectory from biased dynamics and use the TMDmap to approximate committor, comparing the dynamics at various temperatures.

\paragraph{Alanine dipeptide in vacuum}

Alanine dipeptide  ($CH_3-CO-NH-C_{\alpha}HCH_3-CO-NH-CH_3$) is commonly used as a toy problem for sampling studies since it has two well-defined metastable states which can be parameterized by the dihedral angles $\phi$ between $C, N, C_{\alpha}, C$ and $\psi$ defined between $N, C_{\alpha}, C, N$. 

We simulate a $20$ nanosecond  Langevin dynamics trajectory (using the BAOAB integrator~\cite{leimkuhler2015computation}) at temperature $300K$ with a 2fs stepsize, friction $\gamma=1{\rm ps}^{-1}$ and periodic boundary conditions using the openmmtools~\cite{openmmtools} library where the alanine dipeptide in vacuum is provided by the AMBER ff96 force field. 
We sub-sample and RMSD-align the configurations with respect to a reference one leaving only $5\times10^4$ points for the diffusion map analysis. We use kernel~\eqref{eq:kernel} with $\e=1$ and the Euclidean metric.

\begin{figure}
    \centering
    \begin{subfigure}[b]{0.48\textwidth}
        \includegraphics[width=\textwidth]{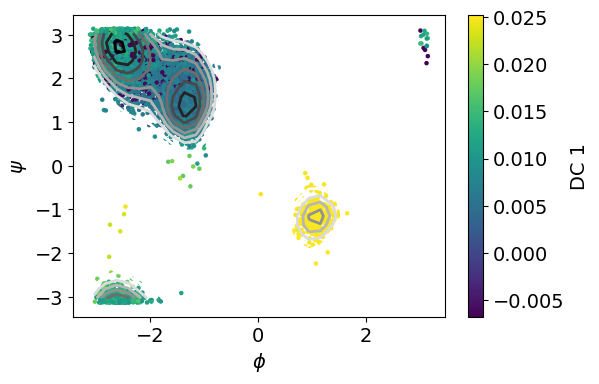}
    \end{subfigure}
   ~
    \begin{subfigure}[b]{0.45\textwidth}
        \includegraphics[width=\textwidth]{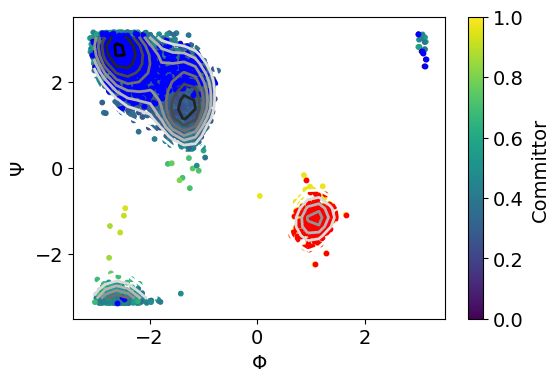}
    \end{subfigure}
 
\caption{Alanine dipeptide.  (a) The two metastable states uncovered by the first diffusion coordinate. (b) The committor function over sets A (red) and B (blue).  The first eigenvector has a very high correlation with the committor function.}
 \label{dominant_cv}
 \end{figure}
 
\begin{figure}
\centering
\begin{subfigure}[b]{0.45\textwidth}
        \includegraphics[width=\textwidth]{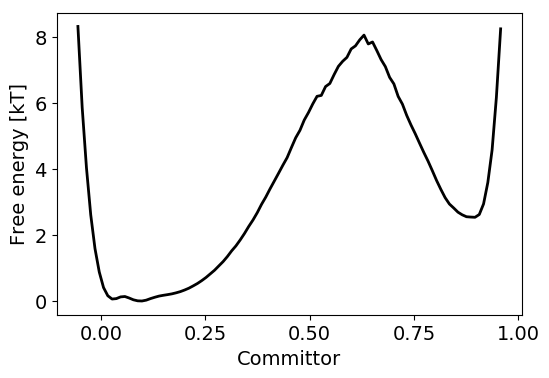}
        
         \end{subfigure}
         \caption{Extrapolated free energy profile along the committor function for alanine dipeptide.}
 
 \label{alanine_all__committor_fe_ext}
\end{figure}

 In  Figure~\ref{dominant_cv}(a)  we show the first diffusion coordinate, which has opposite signs on the two metastable states. Since the transition is very rare, there are only very few points in the vicinity of the saddle point, the dominant eigenvector however clearly parameterizes the dihedral angles, and separates the two metastable sets.
 In order to define the reactive sets, we use the first diffusion coordinate as described in Section~\ref{committors}.  Figure~\ref{dominant_cv} (a) shows the dominant diffusion coordinate, whereas the  Figure~\ref{dominant_cv}(b)  shows the committor function. We observe that the committor strongly correlates with the first eigenvector, which is expected due to the definition of the metastable state given by the dominant eigenvector. Figure~\ref{alanine_all__committor_fe_ext} shows the free energy profile of the committor function, which was extended from the sub-sampled points to the trajectory of length $20$ nanoseconds using nonlinear regression (a multi-layer perceptron)\footnote{More precisely, the subset served as a training set and fitted model was evaluated on the full trajectory. We have performed cross-validation to find the right parameters of the neural network.}.

\begin{figure}

    \centering
    \begin{subfigure}[b]{0.5\textwidth}
        \includegraphics[width=\textwidth]{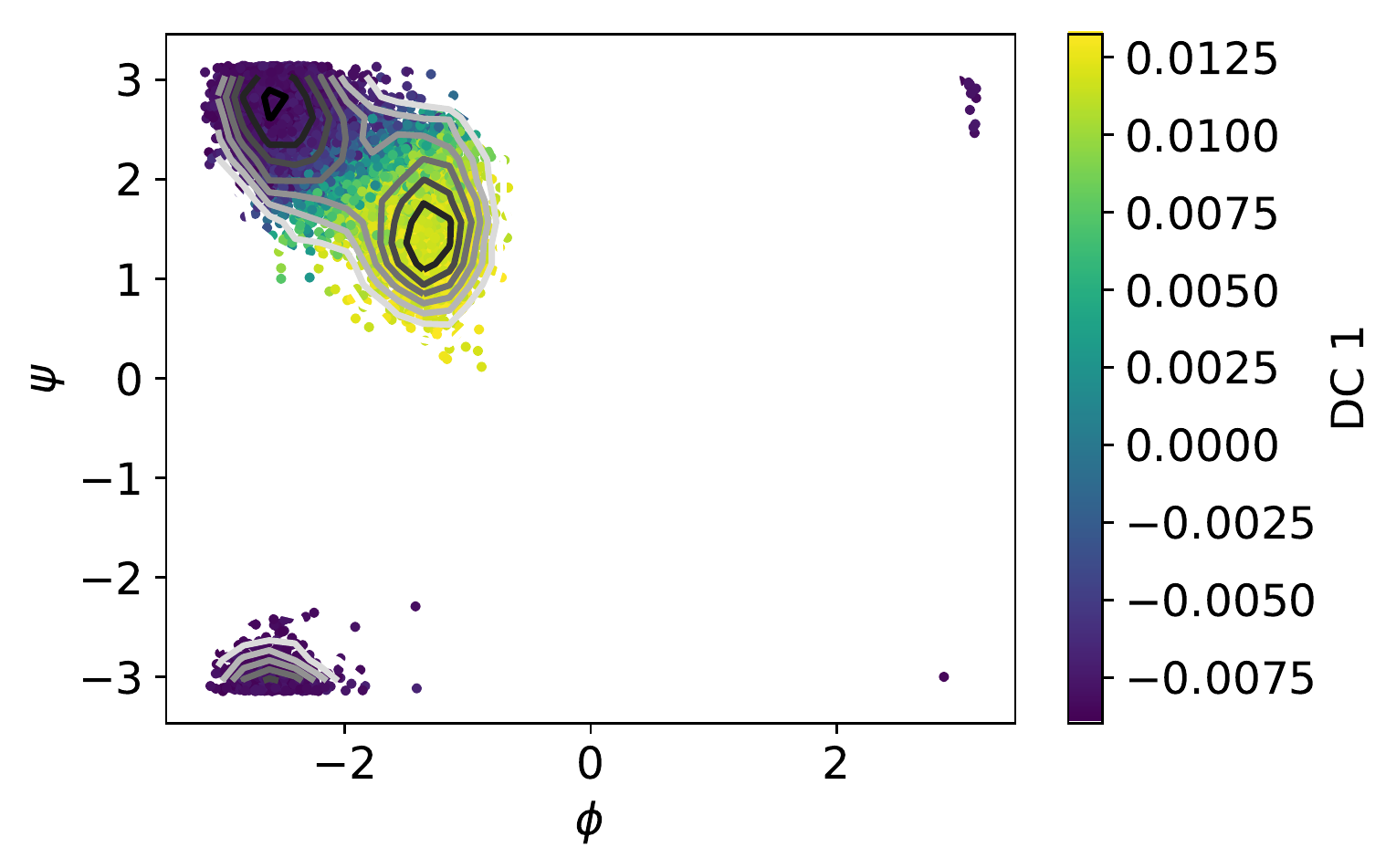}
         \end{subfigure}
   ~
    \begin{subfigure}[b]{0.45\textwidth}
        \includegraphics[width=\textwidth]{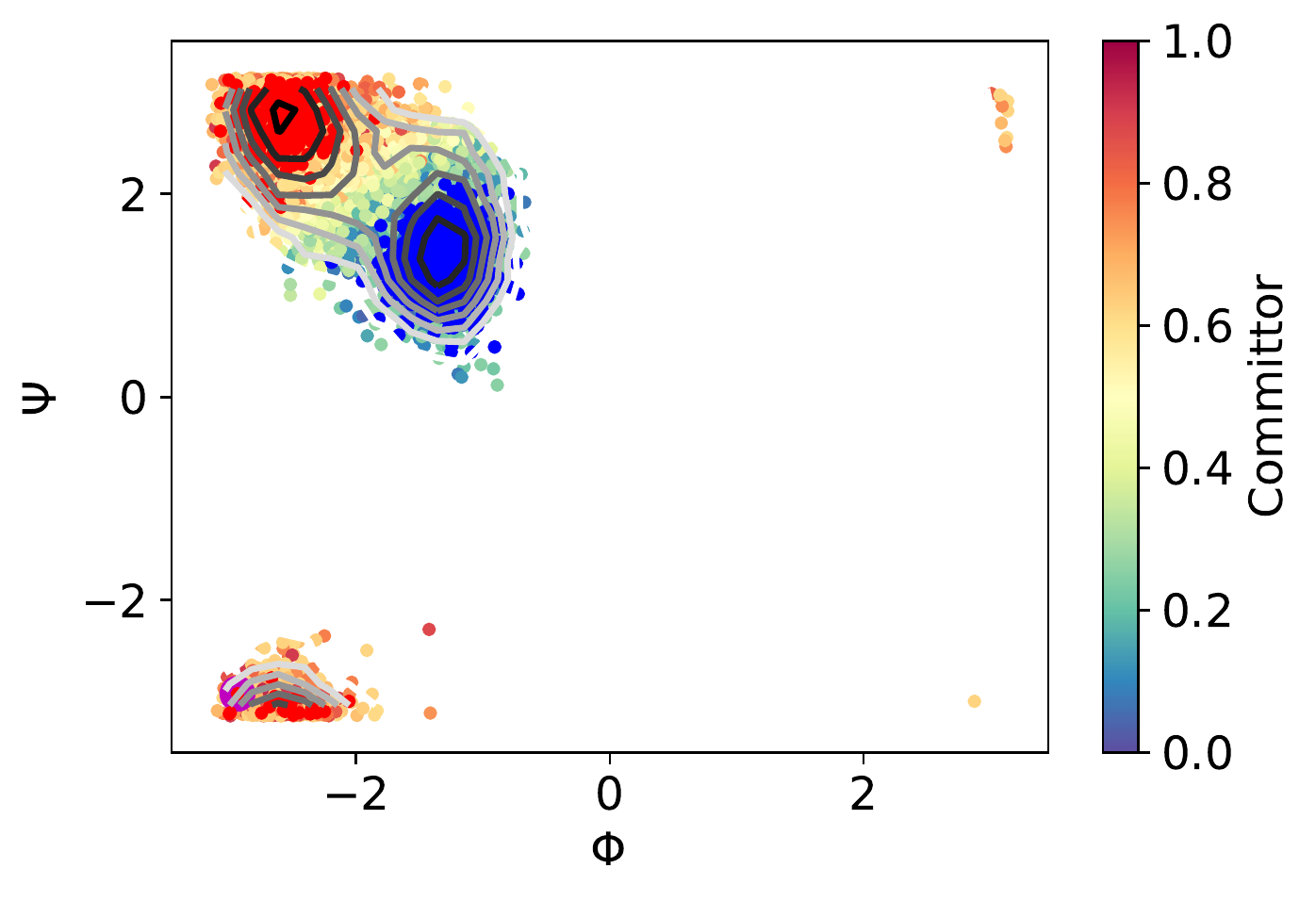}
         \end{subfigure}
         
 \caption{The first metastable state of alanine dipeptide. (a) The dominant eigenvector parameterizes the two wells of the first metastable state. (b) The committor approximation with sets $A, B$ in blue and red.}
 \label{alanine_first_state_dominant_cv}
\end{figure}

\begin{figure}
\centering
\begin{subfigure}[b]{0.45\textwidth}
        \includegraphics[width=\textwidth]{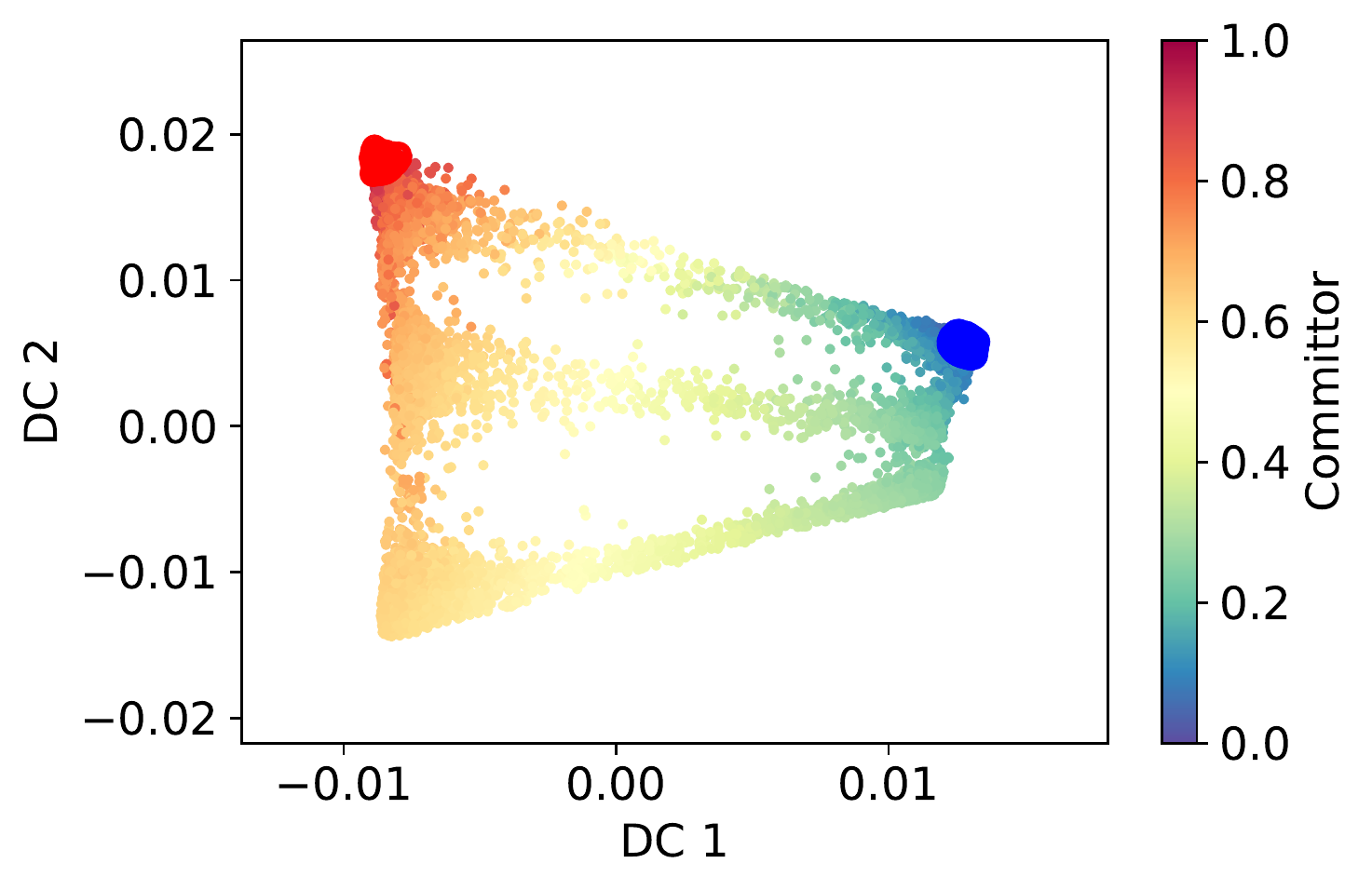}
        
         \end{subfigure}
         \caption{The embedding in the first two diffusion coordinates, coloured by the committor values. The sets $A, B$ are showed in blue and red respectively. }
 
 \label{alanine_first_state_dominant_cv_dc}
\end{figure}

In the previous example, we have used a globally converged trajectory.  The diffusion map analysis therefore describes the dynamics from the global perspective. In order to illustrate the local perspective, we compute the diffusion maps and the committor approximation from a trajectory which has not left the first metastable state. The first eigenvector parameterizes the two wells of this metastable state and we define the reactive states as before using the first DC. The approximated committor assigns the probability $0.5$ correctly to the saddle point between the two wells. Figure~\ref{alanine_first_state_dominant_cv}(a) shows the dominant eigenvector, Figure~\ref{alanine_first_state_dominant_cv}(b) depicts the committor approximation with the automatically chosen sets. Figure~\ref{alanine_first_state_dominant_cv_dc} shows the corresponding diffusion map embedding coloured with respect to the committor values. Note that the diffusion maps used on the samples from this metastable state, whose distribution is the QSD, correlate with the dihedral angles. This observation suggests that alanine dipeptide in vacuum is a trivial example for testing enhanced dynamics methods using collective variables learned on-the-fly because it is likely that the slow dynamics of the metastable state are similar to the global slow dynamics, which might not be the case in more complicated molecules.



\paragraph{Deca-alanine}

\begin{figure}
    \centering
    \begin{subfigure}[b]{0.45\textwidth}  
        \includegraphics[width=\textwidth]{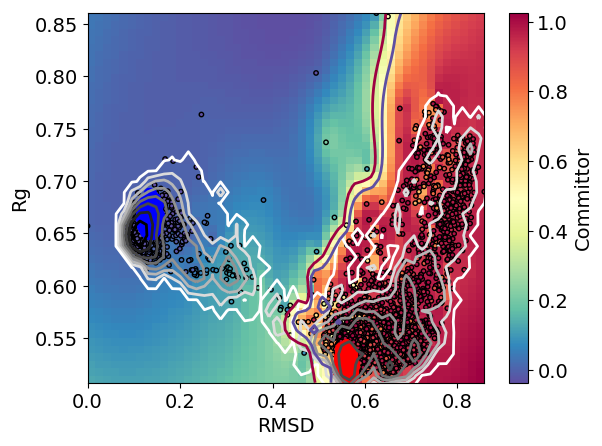}
                         \caption{300K}
         \end{subfigure}
   ~
    \begin{subfigure}[b]{0.45\textwidth}
        \includegraphics[width=\textwidth]{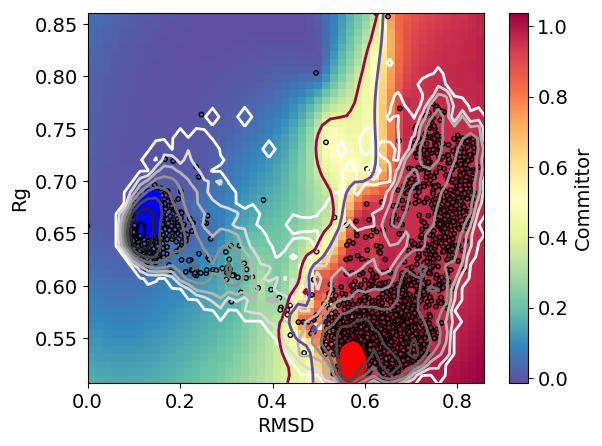}
                         \caption{354K}
         \end{subfigure}

        ~
          \begin{subfigure}[b]{0.45\textwidth}
        \includegraphics[width=\textwidth]{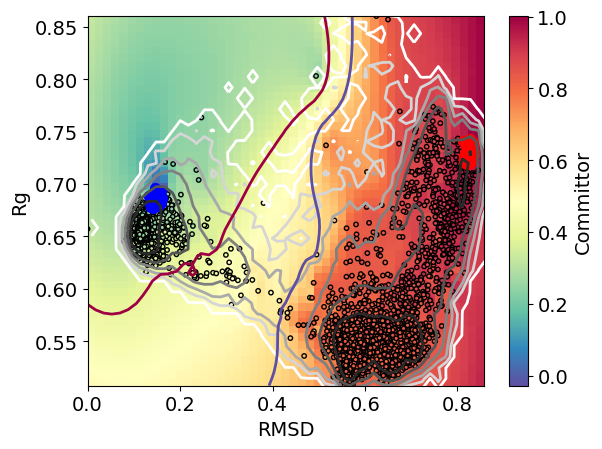}
                         \caption{413K}
         \end{subfigure}
   ~
    \begin{subfigure}[b]{0.45\textwidth}
        \includegraphics[width=\textwidth]{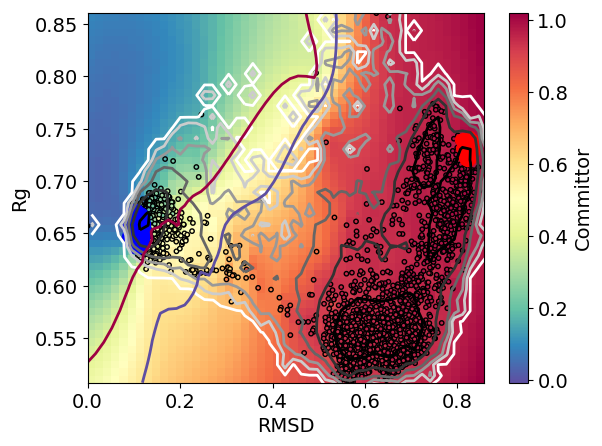}
            \caption{485K}
         \end{subfigure}  
        \caption{Committor of deca-alanine. Sets A(blue) and B1 (red, top) and B2 (red, bottom) were obtained from the first DC. The transition region is showed by committor isosurface lines at $[0.4, 0.6]$. The grey-scale contours show the free-energies. Note that the slowest dynamics is different between low temperatures ($300K$ and $345K$) and high temperatures ($413K$ and $485K$).} 
 \label{alanine12_fe}
\end{figure}
\begin{figure}
    \centering
    \begin{subfigure}[b]{0.45\textwidth}
        \includegraphics[width=\textwidth]{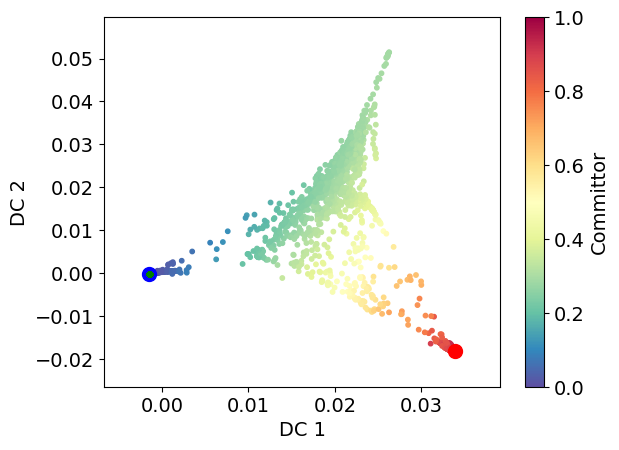}
                         \caption{300K}
         \end{subfigure}
   ~
    \begin{subfigure}[b]{0.45\textwidth}
        \includegraphics[width=\textwidth]{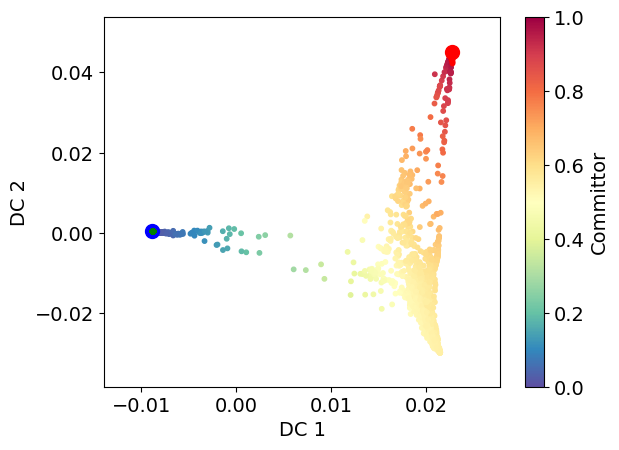}
                         \caption{354K}
         \end{subfigure}

        ~
          \begin{subfigure}[b]{0.45\textwidth}
        \includegraphics[width=\textwidth]{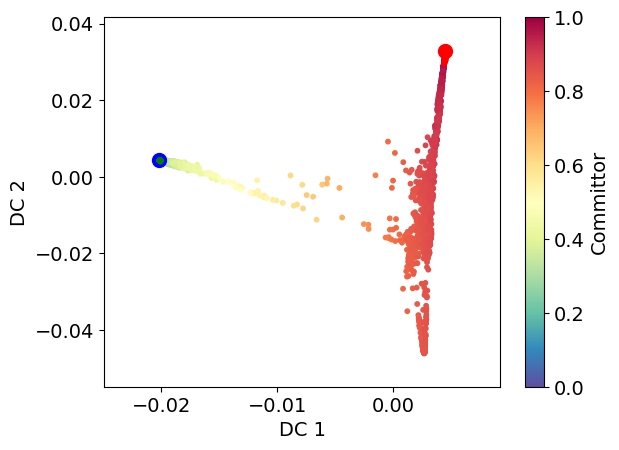}
                         \caption{413K}
         \end{subfigure}
   ~
    \begin{subfigure}[b]{0.45\textwidth}
        \includegraphics[width=\textwidth]{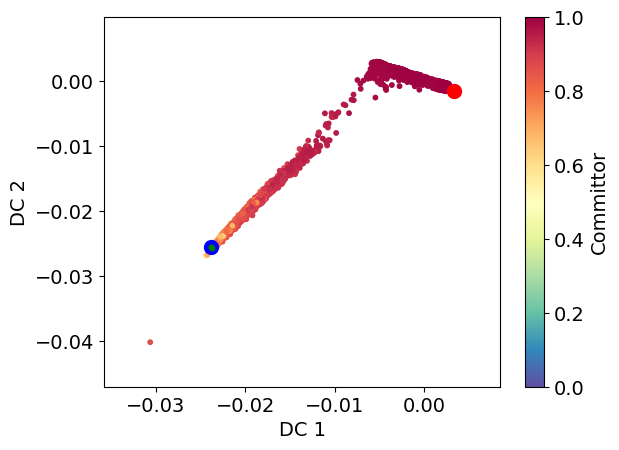}
            \caption{485K}
         \end{subfigure}  
        \caption{Committor of deca-alanine in diffusion coordinates.} 
 \label{alanine12_dc}
\end{figure}

In the following example, we use a long trajectory of the deca-alanine molecule. This molecule has $132$ atoms and a very complex free-energy landscape. A kinetic analysis was done in~\cite{Vitalini2015} showing the force-field-dependent slowest dynamics processes. A representation of the molecular conformations associated to the two main metastable states are shown in Figure~\ref{alanine12}. The system is highly metastable: a standard simulation at $300$K requires at least $5\mu$s to converge~\cite{Vitalini2015}. Therefore, the trajectory studied here was obtained using the infinite-swap simulated tempering (ISST)~\cite{martinsson2018simulated} method with temperatures in the range from $300$K to $500$K (nominal stepsize $2$fs and nominal simulation length $2\mu s$). This method is incorporated into the MIST~\cite{bethune2018mist} library which is coupled to the GROMACS and Amber96 forcefields. ISST provides the weights which are necessary to recover the Boltzmann distribution at a given temperature. We use the weights within TMDmap allowing us to efficiently compute the committor iso-surfaces at various temperatures. We extend the committors in the root-mean-square deviation (RMSD) and the radius of gyration (Rg) for better visualisation in Figure~\ref{alanine12_fe} (however, note that the diffusion maps were applied to all degrees of freedom). We automatically identify the metastable sets using the dominant eigenvector. Note that at lower temperatures of $300$K and $354$K, the two states are $A$ and $B1$, while at higher temperatures they are $A$ and $B2$. See Figure~\ref{alanine12} for the corresponding conformations. The reason is that at the lower temperature, the dominant barrier is enthalpic and at higher temperatures, it is rather entropic, suggesting that the slowest transition is between the states $A$ and $B2$, a state which is not very probable at low temperatures. The different dynamical behaviours can be also seen by the varying $0.5$-committor probability shown in Figures~\ref{alanine12_fe}. The same result is also shown in the space of the first two diffusion map coordinates in Figures~\ref{alanine12_dc}.  

\section{From local to global: defining metastable states and enhanced sampling}
\label{sec: accelerated sampling}

In this section, we first illustrate how the spectrum computed from diffusion maps converges within the quasi-stationary distribution. Next, we use diffusion coordinates built from samples within the metastable state to identify local CVs, which are physically interpretable and valid over the whole state space. Finally, we demonstrate that when used in combination with enhaced dynamics as metadynamics, these CVs lead to more efficient global exploration.   The combination of these three procedures defines an algorithm for enhanced sampling which we formalize and briefly illustrate.

\subsection{On-the-fly identification of metastable states}

 The local perspective introduced in Section \ref{section qsd} allows us to define a 
metastable state as an ensemble of configurations (snapshots) along a trajectory, 
for which the diffusion map spectrum converges. The idea is that, when 
trapped in a metastable state, one can compute the spectrum of the 
infinitesimal generator associated with the QSD, which will typically change when going to a new
metastable state.

\begin{figure}
    \centering
    \begin{subfigure}[b]{0.45\textwidth}
        \includegraphics[width=\textwidth]{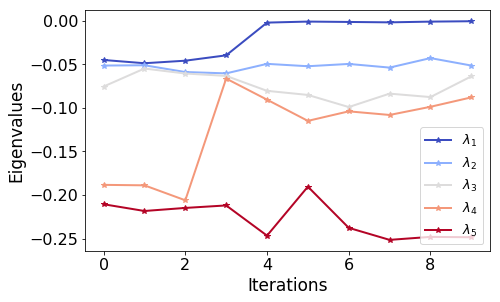}
    \end{subfigure}
    ~
    \begin{subfigure}[b]{0.45\textwidth}
        \includegraphics[width=\textwidth]{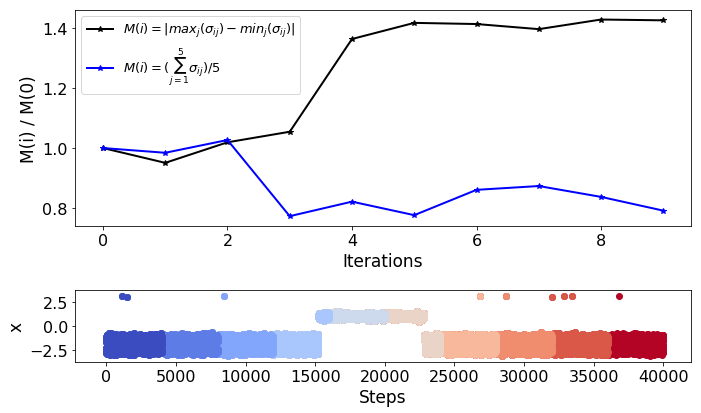}
    \end{subfigure}
   \caption{The first dominant eigenvalues of
     matrix $L_{\e}$ with fixed number of points from alanine dipeptide trajectory. The bottom right plot
     shows the values of $\phi$-angle over the simulation steps, with various colors corresponding to the iterations after which the spectrum is recomputed. Above, we plot the two functions of the dominant eigenvalues, computed over each iteration, which indicate the exit from the state as the eigenvalues suddenly change.  Before the first
     transition to the second state, the spectrum is the one of
     $\Lop_{\Omega}$ (see~\eqref{eq: qsd operator} with $\alpha=\frac
     1 2$), where $\Omega$ is the metastable state where the process
     is trapped. After
     the transition, the spectrum is evolving towards the spectrum of the
     operator $\Lop_{\mathbb R^d}=-\beta \nabla V + \Delta$ on the whole domain.   }
  \label{fig: alanine_iterations_eigenvalues}      
\end{figure}

To illustrate this, we analyze an alanine dipeptide trajectory.
Every $4000$ steps, we compute the
first dominant eigenvalues of the diffusion map matrix $L_{\e}$
(with $\alpha=1/2$) by sampling $2000$ points from the trajectory $\{x_n\}^m_{n=0}$ until the end of the last iteration $\tau=m\de t$.  We observe in  Figure~\ref{fig: alanine_iterations_eigenvalues}(left), that the sampling has locally equilibrated during the first 3 iterations within the metastable state, and the eigenvalues have converged.
The transition occurs after the fourth iteration, which we can see in Figure~\ref{fig: alanine_iterations_eigenvalues}(bottom right) which shows the values of the dihedral angle $\phi$ during the simulation step. In the left figure, we  clearly observe a change in the spectrum at this point, with an increase in the spectral gap. After the trajectory has exited the metastable state, the eigenvalues
begin evolving to new values, corresponding to the spectrum of the
operator on the whole domain.
The change in the spectrum allows us to detect the exit from the metastable state. Instead of tracking each of the first eigenvalues separately, we compute the average and the maximal difference of the dominant eigenvalues, the values of which we plot in Figure~\ref{fig: alanine_iterations_eigenvalues}(top right).



This example illustrates the local perspective on the diffusion maps:
their application on a partially explored distribution provides an
approximation of a different operator since the used samples are
distributed with respect to a quasi-stationary distribution. It is therefore possible to use
samples from the local equilibrium to learn the slowest dynamics
 within the metastable state. 

This observation can be used together with biasing techniques 
to improve sampling. 
 
\subsection{Enhanced sampling procedure for complex molecular systems}

 
 We now describe an outline of an algorithm for enhanced sampling.

 Sampling from the QSD allows us
to build high quality local CVs (within the meastable state) by looking for the most correlated physical CVs to the DCs. In typical practice the CVs are chosen from a list of physically relevant candidates, as for example the backbone dihedral angles or atom-atom contacts. There are several advantages to using physical coordinates instead of the abstract DC.  First, the artificial DCs are only defined on the visited states, i.e. inside the metastable state. Extrapolated DCs outside the visited state lose their validity further from points used for the computation.  By contrast, the physical coordinates can typically be defined over the entire state space. Second, the slowest physical coordinates might provide more understanding of the metastable state.

Once the best local CVs have been identified, 
we can use metadynamics to enhance the sampling, effectively driving the dynamics to exit the metastable state. In the next iteration, we suggest to use TMDmap to unbias the influence of metadynamics on the newly generated trajectory.

The above strategy can be summarized in the following algorithm (Enhanced sampling algorithm based on local-global analysis):
\begin{enumerate}
\item Run molecular dynamics until the spectrum is converged (the reference  walker is trapped)
\item {\em [Optional]} Refine the convergence of the spectrum and the DC by using
a Fleming-Viot~\cite{binder-simpson-lelievre-15} process within the metastable state (each replica being initialized with the whole 
initial trajectory of the reference dynamics, for the diffusion map computation)
\item
Identify CVs which are the most correlated with DCs
\item Build an effective bias using, for example,  metadynamics based on these CVs
\item When an exit is observed for the reference walker (through a change 
in the spectrum), restart the procedure from step 1, keeping in memory 
the bias built using TMDmap
\end{enumerate}
 
As proof-of-concept, we will illustrate this method in a simplified setup based on metadynamics in the case of the alanine dipeptide. We stress that the construction of sampling procedures for general macromolecules is a complex undertaking and many challenges still need to be addressed to implement the full proposed methodology for larger systems; this is the subject of current study by the authors.
 
\begin{figure}
     \centering
        \includegraphics[width=0.5\textwidth]{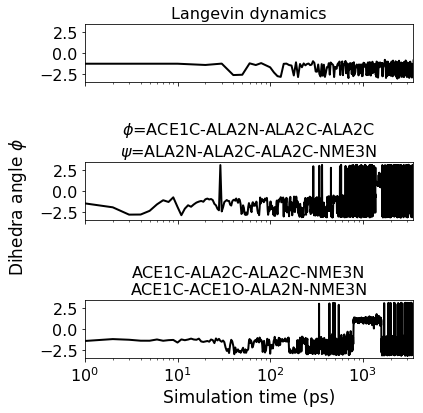}
    \caption{Alanine dipeptide: Metadyamics with adaptive-CV (bottom)  angles outperforms Langevin dynamics (top). Metadynamics with $\phi, \psi$ angles (middle) is also more efficient than standard Langevin. 
    }
 \label{fig:metadynamics}
\end{figure}

 We first run a short trajectory to obtain samples from the first metastable state\footnote{The 'first' state corresponds to the left top double well in Figure~\ref{dominant_cv}.}. This sampling is done by Fleming-Viot process \cite{binder-simpson-lelievre-15} with boundary defined by the converged spectrum which we regularly recompute during the sampling\footnote{Alternatively, one can define the boundary  by the free-energy in $\phi, \psi$ angles. We also found that resampling using Fleming-Viot improves the quality of diffusion maps.}.  
 We compute the two dominant DCs
 and use them to select the two most correlated from a list of physical coordinates (a similar approach was presented in \cite{ferguson2011integrating}). In this case, we looked at the collection of all possible dihedral angles from all the atoms except  the hydrogens. In order to select the two most correlated CVs we employed the Pearson correlation coefficient,
 \[
\rho(X, Y) = \frac{{\rm cov} (X, Y)}{\sigma_X\sigma_Y},
\]
where $\sigma_Z$ is the standard deviation of a random variable $Z$. 
We found dihedral angles ACE1C-ALA2C-ALA2C-NME3N and ACE1C-ACE1O-ALA2N-NME3N to be the best candidates with highest correlations with the first and the second eigenvector respectively. Note however, that there were also several other CVs with high correlations, among these the $\phi$ and $\psi$ angles. In the next step, we use the most correlated physical CVs within metadynamics and track the $\phi$-angle as a function of time.  Moreover, we  run metadynamics with a priori chosen CV's based on expert knowledge, specifically the $\phi, \psi$ angles.  As shown in Figure~\ref{fig:metadynamics}, there is no transition for the standard dynamics (top).  On the other hand, metadynamics with the learned CVs exhibits several transitions, see Figure~\ref{fig:metadynamics}(bottom), similarly to the knowledge based CVs (middle). We have also compared the least correlated CVs for metadynamics and this approach was much worse than Langevin dynamics: we did not observe any exit during the expected Langevin exit time. These numerical experiments are strongly suggestive of the relevance of the most correlated CVs with the dominant DCs\footnote{In this example, we did not compute the full expected exit times since the difference was significant for several realizations. To estimate the actual speed up of the sampling method, one would need to run thousands of trajectories and perform a detailed statistical analysis on the results, which is proposed for future study. 
}.  In conclusion, learning the slowest CVs from a local state provides important information allowing to escape the metastable state and hence enhance the sampling. Once the process leaves the first visited metastable state, one keeps the bias, new CVs are computed to update the bias once trapped elsewhere. This process can be iterated, and the weights from metadynamics can subsequently be unbiased by TMDmap to compute the relevant physical coordinates at every iteration.

We have thus demonstrated, at least in this specific case, how the local perspective can be used together with biasing techniques (metadynamics)
to get more rapid sampling of the target distribution.


\section{Conclusion and future work}

Diffusion maps are an effective tool for uncovering a natural collective variable as needed for enhanced sampling. In this work, we have formalized the use of diffusion maps within a metastable state, which provides insight into diffusion map-driven sampling based on iterative procedures.  The main theoretical tool for stating an analytical form of the approximated operator is the quasi-stationary distribution.  This local equilibrium guarantees the convergence of the diffusion map within the metastable state. We have also demonstrated that diffusion-maps, especially the TMDmap, can be used for committor computations in high dimensions. The low computational complexity aids in the analysis of molecular trajectories and helps to unravel the dynamical behaviour at various temperatures. 

We have used the local perspective to identify the metastable state as a collection of states for which the spectrum computed by diffusion maps converges. We use the diffusion map eigenfunctions to learn physical coordinates which correspond to the slowest modes of the metastable state. This information not only helps to understand the metastable state, but leads to iterative procedure which can enhance the sampling.


Following the encouraging results we obtained in the last section, other techniques can be explored to fuel the iterative diffusion map sampling: for example the adaptive biasing force method~\cite{darve2008adaptive}, metadynamics, or dynamics biasing techniques as adaptive multilevel splitting~\cite{cerou2007adaptive}, Forward Flux Sampling~\cite{allen2009forward}. In the case of AMS, committor can be used as the one dimensional reaction coordinate. It will be worth exploring these strategies for more complex molecules.

Finally, as a future work, we point out that the definition of the metastable states using the spectrum computed by diffusion map could be used within an accelerated dynamics algorithm, namely the parallel replica algorithm \cite{perez-uberuaga-voter-15, voter-98}. 
Specifically, one could use the 
Flemming-Viot particle process within the state (Step 2 of Algorithm 1) to estimate the correlation time, using the Gelman-Rubin convergence diagnostic (see \cite{binder-simpson-lelievre-15}), 
restarting the sampling whenever the reference walker leaves prior to convergence and otherwise using trajectories generated from the FV process
to compute exit times (using the diffusion map spectra to identify exit). 


\enlargethispage{20pt}

\section{Data accessibility}
{Data is accessible as "Martinsson, Anton; Trstanova, Zofia. (2019). ISST Deca-Alanine. University of Edinburgh. School of Mathematics.https://doi.org/10.7488/ds/2491". 
The following libraries were used:
- pydiffmap
https://github.com/DiffusionMapsAcademics/pyDiffMap
- OpenMM
Eastman, Peter, et al. "OpenMM 7: Rapid development of high-performance algorithms for
molecular dynamics." PLoS computational biology 13.7 (2017): e1005659
}

\section{Acknowledments}
{B.L and Z.T. were supported by EPSRC Grant
  EP/P006175/1. B.L. was further supported by the Alan Turing
  Institute (EPSRC EP/N510129/1) as a Turing Fellow. T.L. is supported by the European Research Council under the European Union's Seventh Framework Programme (FP/2007-2013) / ERC Grant Agreement number 614492.}

{The authors would like to thank Ben Goddard and Antonia Mey (both at University of Edinburgh) for helpful discussions. }

\bibliographystyle{plain}

\end{document}